\documentclass[12pt]{article}
\usepackage{graphicx,amssymb,amsfonts,amsmath,epsfig}

\makeatletter
\DeclareSymbolFont{AMSb}{U}{msb}{m}{n}
\DeclareSymbolFontAlphabet{\mathbb}{AMSb}
\renewcommand{\section}{\@startsection{section}{1}{\z@}%
                                    {-7ex \@plus -1ex \@minus -.2ex}%
                                    {2.5ex \@plus.2ex}%
                                    {\normalfont\large\scshape\centering}}
\renewcommand{\subsection}{\@startsection{subsection}{2}{\z@}%
                                       {-5ex \@plus -1ex \@minus -.2ex}%
                                       {1.5ex \@plus.2ex}%
                                       {\normalfont\normalsize\scshape}}
\renewcommand{\subsubsection}{\@startsection{subsubsection}{3}{\z@}%
                                       {-5ex \@plus -1ex \@minus -.2ex}%
                                       {1.5ex \@plus.2ex}%
                                       {\normalfont\normalsize\scshape}}

\renewcommand\@seccntformat[1]{\ignorespaces\csname #1name\endcsname\space
                               \csname the#1\endcsname.\quad}   
\newdimen\captionmargin
\newdimen\captionindent
\newdimen\captionwidth
\newcommand{\captionfont}{\slshape}
\newcommand\@captionlabel[1]{\textsc{#1:}\space}
\long\def\@makecaption#1#2{%
  \vskip\abovecaptionskip
  \captionwidth\hsize
  \advance\captionwidth -2\captionmargin
  \sbox\@tempboxa{\@captionlabel{#1}\captionfont #2}%
  \ifdim \wd\@tempboxa >\captionwidth
    \ifdim\captionindent>\z@
      \advance\captionwidth -\captionindent
      \hskip\captionindent
    \fi
    \hskip\captionmargin
    \parbox[t]{\captionwidth}{\leavevmode\hskip-\captionindent
      \@captionlabel{#1}\captionfont #2}%
  \else
    \global \@minipagefalse
    \hb@xt@\hsize{\hfil\box\@tempboxa\hfil}%
  \fi
  \vskip\belowcaptionskip}
\def\eqnarray{%
   \stepcounter{equation}%
   \def\@currentlabel{\p@equation\theequation}%
   \global\@eqnswtrue
   \m@th
   \global\@eqcnt\z@
   \tabskip\@centering
   \let\\\@eqncr
   $$\everycr{}\halign to\displaywidth\bgroup
       \hskip\@centering$\displaystyle\tabskip\z@skip{##}$\@eqnsel
      &\global\@eqcnt\@ne$\;\hfil{##}$\hfil
      &\global\@eqcnt\tw@$\;\displaystyle{##}$\hfil\tabskip\@centering
      &\global\@eqcnt\thr@@ \hb@xt@\z@\bgroup\hss##\egroup
         \tabskip\z@skip
      \cr}
\ifcase \@ptsize
\or
\or
\fi
  \divide\@tempdima\baselineskip
  \@tempcnta=\@tempdima
\makeatother
\begin{document}

%
%

\renewcommand{\theequation}{\arabic{section}.\arabic{equation}}
\renewcommand{\thefigure}{\arabic{figure}}
\newcommand{\gapprox}{%
\mathrel{%
\setbox0=\hbox{$>$}\raise0.6ex\copy0\kern-\wd0\lower0.65ex\hbox{$\sim$}}}
\textwidth 165mm \textheight 220mm \topmargin 0pt \oddsidemargin 2mm
\def\ib{{\bar \imath}}
\def\jb{{\bar \jmath}}

\newcommand{\ft}[2]{{\textstyle\frac{#1}{#2}}}
\newcommand{\be}{\begin{equation}}
\newcommand{\ee}{\end{equation}}
\newcommand{\bea}{\begin{eqnarray}}
\newcommand{\eea}{\end{eqnarray}}
\newcommand{\cx}{\overset{\circ}{x}_2}
\def\CN{$\mathcal{N}$}
\def\CH{$\mathcal{H}$}
\def\hg{\hat{g}}
\newcommand{\bref}[1]{(\ref{#1})}
\def\espai{\;\;\;\;\;\;}
\def\zespai{\;\;\;\;}
\def\avall{\vspace{0.5cm}}
\newtheorem{theorem}{Theorem}
\newtheorem{acknowledgement}{Acknowledgement}
\newtheorem{algorithm}{Algorithm}
\newtheorem{axiom}{Axiom}
\newtheorem{case}{Case}
\newtheorem{claim}{Claim}
\newtheorem{conclusion}{Conclusion}
\newtheorem{condition}{Condition}
\newtheorem{conjecture}{Conjecture}
\newtheorem{corollary}{Corollary}
\newtheorem{criterion}{Criterion}
\newtheorem{defi}{Definition}
\newtheorem{example}{Example}
\newtheorem{exercise}{Exercise}
\newtheorem{lemma}{Lemma}
\newtheorem{notation}{Notation}
\newtheorem{problem}{Problem}
\newtheorem{prop}{Proposition}
\newtheorem{rem}{{\it Remark}}
\newtheorem{solution}{Solution}
\newtheorem{summary}{Summary}
\numberwithin{equation}{section}
\newenvironment{pf}[1][Proof]{\noindent{\it {#1.}} }{\ \rule{0.5em}{0.5em}}
\newenvironment{ex}[1][Example]{\noindent{\it {#1.}}}

\makeatletter
\def\flE{\begin{picture}(0,0)
   \put( 0.25,    0){\vector( 1, 0){0.50}}
   \@ifstar{\@flE}{\@@flE}}
\def\@flE  #1{\put( 0.5 ,-0.03){\makebox(0,0)[ t]{$#1$}}\end{picture}}
\def\@@flE #1{\put( 0.5 , 0.03){\makebox(0,0)[ b]{$#1$}}\end{picture}}
\def\flNE{\begin{picture}(0,0)
   \put( 0.18, 0.18){\vector( 1, 1){0.64}}
   \@ifstar{\@flNE}{\@@flNE}}
\def\@flNE #1{\put( 0.52, 0.48){\makebox(0,0)[tl]{$#1$}}\end{picture}}
\def\@@flNE#1{\put( 0.48, 0.52){\makebox(0,0)[br]{$#1$}}\end{picture}}
\def\flN{\begin{picture}(0,0)
   \put(    0, 0.20){\vector( 0, 1){0.60}}
   \@ifstar{\@flN}{\@@flN}}
\def\@flN  #1{\put( 0.03, 0.5 ){\makebox(0,0)[ l]{$#1$}}\end{picture}}
\def\@@flN #1{\put(-0.03, 0.5 ){\makebox(0,0)[ r]{$#1$}}\end{picture}}
\def\flNW{\begin{picture}(0,0)
   \put(-0.18, 0.18){\vector(-1, 1){0.64}}
   \@ifstar{\@flNW}{\@@flNW}}
\def\@flNW #1{\put(-0.48, 0.52){\makebox(0,0)[bl]{$#1$}}\end{picture}}
\def\@@flNW#1{\put(-0.52, 0.48){\makebox(0,0)[tr]{$#1$}}\end{picture}}
\def\flW{\begin{picture}(0,0)
   \put(-0.25,    0){\vector(-1, 0){0.50}}
   \@ifstar{\@flW}{\@@flW}}
\def\@flW  #1{\put(-0.5 , 0.03){\makebox(0,0)[ b]{$#1$}}\end{picture}}
\def\@@flW #1{\put(-0.5 ,-0.03){\makebox(0,0)[ t]{$#1$}}\end{picture}}
\def\flSW{\begin{picture}(0,0)
   \put(-0.18,-0.18){\vector(-1,-1){0.64}}
   \@ifstar{\@flSW}{\@@flSW}}
\def\@flSW #1{\put(-0.52,-0.48){\makebox(0,0)[br]{$#1$}}\end{picture}}
\def\@@flSW#1{\put(-0.48,-0.52){\makebox(0,0)[tl]{$#1$}}\end{picture}}
\def\flS{\begin{picture}(0,0)
   \put(    0,-0.2 ){\vector( 0,-1){0.60}}
   \@ifstar{\@flS}{\@@flS}}
\def\@flS  #1{\put(-0.03,-0.5 ){\makebox(0,0)[ r]{$#1$}}\end{picture}}
\def\@@flS #1{\put( 0.03,-0.5 ){\makebox(0,0)[ l]{$#1$}}\end{picture}}
\def\flSE{\begin{picture}(0,0)
   \put( 0.18,-0.18){\vector( 1,-1){0.64}}
   \@ifstar{\@flSE}{\@@flSE}}
\def\@flSE #1{\put( 0.48,-0.52){\makebox(0,0)[tr]{$#1$}}\end{picture}}
\def\@@flSE#1{\put( 0.52,-0.48){\makebox(0,0)[bl]{$#1$}}\end{picture}}
\def\capsa(#1,#2)#3{\put(#1,#2){\makebox(0,0){$#3$}}}
\def\indiag{\@ifnextchar [{\@indiag}{\@indiag[15ex]}}
\def\@indiag[#1](#2,#3){\begingroup
   \setlength{\unitlength}{#1}
   \medskip
   \begin{center}
   \begin{picture}(#2,#3)}
\def\exdiag{\end{picture}
   \end{center}
   \medskip
   \endgroup}
\makeatother

\thispagestyle{empty}

\begin{flushright}\scshape
September 2003\\
\end{flushright}
\vskip1cm

\begin{center}

{\LARGE\scshape Consistent and inconsistent truncations.
Some results and the issue of the correct uplifting of
solutions.
\par}
\vskip15mm

\textsc{Josep M. Pons$^{a}$ and Pere Talavera$^{b}$}
\par\bigskip
$^a${\em
Departament d'Estructura i Constituents de la Mat\`eria,
Universitat de Barcelona,\\
Diagonal 647, E-08028 Barcelona, Spain.}\\[.1cm]
$^b${\em
Departament de F{\'\i}sica i Enginyeria Nuclear,
Universitat Polit\`ecnica de Catalunya,\\
Jordi Girona 1--3, E-08034 Barcelona, Spain.}\\[.1cm]
\vspace{5mm}
\end{center}

\section*{Abstract}
We clarify the existence of two different types of truncations of
the field content in a theory, the consistency of each type being
achieved by different means. A proof is given of the
conditions to have a consistent truncation in the case of
dimensional reductions induced by independent Killing vectors. We
explain in what sense the tracelessness condition found by Scherk
and Schwarz is not only a necessary condition but also a {\it
sufficient} one for a consistent truncation. The reduction of the
gauge group is fully performed showing the existence of a sector
of rigid symmetries. We show that truncations originated by the
introduction of constraints will in general be inconsistent, but
this fact does not prevent the possibility of correct upliftings
of solutions in some cases. The presence of constraints has
dynamical consequences that turn out to play a fundamental
role in the correctness of the uplifting procedure.

\vspace{3mm} \vfill{ \hrule width 5.cm \vskip 2.mm {\small
\noindent E-mail: pons@ecm.ub.es, pere.talavera@upc.es }}

\newpage
\setcounter{page}{1}


\tableofcontents       %
\vskip 1cm             %

\setcounter{equation}{0}

\section{Introduction}
\label{sec:intro}

In this paper we examine the conditions under which a certain type
of dimensional reductions are consistent truncations of the
theory, and address some issues concerning the elimination of
degrees of freedom, including the possibility of upliftings even in
cases when the truncation is not strictly a consistent one.

Some questions of language must be addressed first. A basic one is
the concept of {\sl consistent truncation}.
We find it convenient to use this concept in a
wide sense, of which a dimensional reduction will just
be an important, but special case. Consider a Lagrangian
${\mathcal L}$ as the starting point, for a certain number of
dimensions of the space-time. We can produce a truncation of it by
essentially two methods --or a mixture of both:
\begin{itemize}
\item[{\it{i)}}] First-type: by reducing the dimension of the
space-time (Kaluza-Klein dimensional reduction)\footnote{We shall
only consider the case when the Killing vectors are independent.}
while keeping unchanged the number of degrees of freedom attached
to every space-time point\footnote{The issue of degrees of freedom
per space-time point is further clarified below.}.
\end{itemize}
\begin{itemize}
\item[{\it{ii})}] Second-type: by introducing constraints that
reduce the number of independent fields --or field components--
defining the theory\footnote{Our considerations will be
restricted to constraints in configuration space.}.
\end{itemize}
These two procedures are usually applied altogether under the
common concept of dimensional reduction, but we think it is very
convenient to maintain a clear distinction between them. In both
cases we are producing a {\sl truncation} in the field content of
the theory, either because Kaluza-Klein modes are eliminated in
the dimensional reduction or because some field components become
redundant due to the presence of constraints. At this point an
issue of {\sl consistency} of such a truncation, be it first-type,
second-type or mixed, arises. Namely, whether the solutions of the
equations of motion (e.o.m.) for ${\mathcal L}_R$ are still
solutions of the e.o.m. for the original ${\mathcal L}$. This
property is expressed graphically as the commutativity of the
following diagram, \indiag(1,1) \capsa(0,0){
\frac{\delta{\mathcal L}}{\delta \Phi}\!=0 } \capsa(0,1){{\mathcal L}}
\capsa(0,1){\flS{{\rm e.o.m.}}} \capsa(0,0){\flE{{\rm Red}.}}
\capsa(0,1){\qquad\quad\flE{{\rm Red}.}} \capsa(1,0){\qquad\
\qquad \qquad\big(\frac{\delta{\mathcal L}}{\delta \Phi}
\big)_{\!_{\!R}}\!=0 \Leftrightarrow \frac{\delta{\mathcal
L}_{\!R}}{\delta \Phi}=0 } \capsa(1.5,1){{\mathcal L}_R}
\capsa(1.5,1){\flS{{\rm e.o.m.}}} \exdiag

\avall

\noindent

A proper definition is the following: {\sl A truncation is said to
be consistent when its implementation at the level of the
variational principle agrees with that at the level of the
equations of motion}, i.e., if both operations commute: first
truncate the Lagrangian and then obtain the equations of motion
(e.o.m.), or first obtain the equations of motion and then
truncate them. This definition of consistent truncation is essentially 
the one appearing in \cite{Duff:1984hn,Duff:jd,Duff:1986ya,Duff:1989cr}
where important clarifications were made concerning the issue of consistency.
We will make at the end of the section some comments on different approaches 
as regards this concept.

Therefore the concept of {\sl consistent truncation} goes much
beyond the idea of dimensional reductions. A complementary aspect
of that of a truncation is that of an \emph{uplifting}, that we
now introduce. To the process of truncation {\sl of theories},
${\mathcal L} \rightarrow{\mathcal L}_R$, going from top to
bottom, there corresponds an opposite process of uplifting {\sl of
solutions}, from bottom to top. The uplifting procedure uses the
same recipes (Killing conditions in first-type truncations,
explicit form of the constraints in the second) that define the
truncation. The basic result in this respect is that {\sl a
consistent truncation guarantees that any solution of the
${\mathcal L}_R$ dynamics can be uplifted to a solution of the
${\mathcal L}$ dynamics}. This result, to be proved in
sections \ref{upl} and \ref{upl2}, reinforces the suitability of the above definition of
a consistent truncation.

\avall

Let us now add some further comments concerning both types of
truncations.

As regards first-type truncations, that is, Kaluza Klein
dimensional reductions (\cite{kaluza,klein}, see \cite{Duff:hr}
and \cite{Appelquist:nr} as general references), let us clearly
state that with the truncation we indeed reduce the Lagrangian and
therefore we are going beyond a pure compactification. In a pure
compactification one uses the topological properties of the
space-time and expands (Fourier modes, Spherical Harmonics, etc.,)
the fields with respect to the ``compactified" coordinates, thus
trading some space-time dimensions for an infinity of states (the
Kaluza-Klein tower of states). It might be that, since the masses
of the Kaluza-Klein modes are inversely proportional to the length
dimension of the compactified structure, only the massless states
play any role in an effective sense, but what we are saying with
the truncation procedure is mathematically different: with the
truncation, the massive KK modes are set to zero, and a new
theory, the truncated theory, is defined (with the same finite
number of degrees of freedom per space-time
point\footnote{Although the dimension of the space-time has been
reduced and therefore the concept of space-time point has been
changed.}).

\avall

One basic result of this paper is a proof of the necessary
and sufficient conditions that dimensional reductions must satisfy
--in the case of independent Killing vector fields-- to yield a
consistent truncation. We think that our proof fills some gap
in the literature, and we expect it will clarify completely the
issue of consistency from the point of view of the equations of
motion derived from variational principles. Previous work in this
subject includes
\cite{hawking;69,Sneddon,Scherk:1979zr,Maccallum:gd,Pons:1998tt}\footnote{
A historic account on the problem of consistent truncations for
Bianchi cosmologies can be found in \cite{bianchi}.}. In this
respect, our work will be a generalisation of \cite{Pons:1998tt},
where, in the context of the Bianchi cosmologies, the reduction to
a mechanical (1-dimensional field theory) case was performed.

It is fundamental for our proof that the Lie algebra of Killing
symmetries be generated by a set of {\sl independent} vector
fields, so that there will be as many space-time dimensions
eliminated as the dimension of the algebra of Killing symmetries.
Many interesting cases of dimensional reduction, including
compactifications on spheres, lie outside this assumption. For
these cases, it seems as of now that there is no general a priori
criterion that could decide, on the basis of the structure of
both the Killing algebra and the space upon which the reduction
takes place, whether a truncation is eventually going to be
consistent, although some necessary conditions are known
\cite{Duff:hr}. At this moment, it still remains,
unsatisfactorily, a matter of heuristics and ansatzs.

Let us also mention that our methods allow for a systematic
study of the reduction of the gauge group. We find in particular
that the reduction of the diffeomorphism group decomposes into
three subgroups, that of diffeomorphisms in the reduced space, a
Yang-Mills gauge group and a linearly realized discrete group of
rigid symmetries.

 \avall

 \avall

 As for
the second-type truncations, driven by the introduction of
constraints, we show that a mechanism parallel to that of
Dirac-Bergmann's theory of constrained system, applies. In
particular we notice that the presence of secondary constraints
--dynamically derived from the original ones-- is a typical
obstruction for the truncation to be consistent. We work out a
simple example of this phenomenon: a dimensional reduction of pure
general relativity (GR)  accompanied with the introduction of constraints that eliminate
the sector of charged scalars\footnote{Charged under the new
Yang-Mills gauge group.}. We show in this particular example that
these --primary-- constraints have in their turn the effect of
introducing new --secondary-- constraints on the Yang-Mills field
strengths. This example allows us to assert that an
uplifting applied to the set of solutions of the truncated theory
that happen to satisfy the secondary constraints is still correct.

Let us give a simple example of a trivial second-type
truncation: the reduction of a theory to its bosonic sector.
Consider a Lagrangian ${\mathcal L}$ describing a theory with
bosons and fermions, the latter fields expressed as odd Grasmann
variables. Since the Lagrangian is an even function, the terms
including fermions will be quadratic at least. Because of that,
setting all fermions to zero is a consistent truncation. We can
therefore consider the truncated Lagrangian ${\mathcal L}_R$ as
the bosonic sector of ${\mathcal L}$, and we are guaranteed that
any --of course bosonic-- solution of ${\mathcal L}_R$ can be
uplifted to a solution of ${\mathcal L}$, that will have all
fermions set to zero.

\avall

As will be shown, a consistent truncation 
guarantees that the uplifting of a solution of the truncated 
theory is always a solution of the untruncated one. But it is 
important to realize that there can exist correct upliftings 
(correct in the sense just given above) even in cases when the 
truncation is not consistent, as long as the candidate solutions 
for the uplifting satisfy the appropriate conditions. In this sense, 
our definition of a consistent truncation through the commutativity 
of the diagram above, although equivalent to defining it by requiring
that all solutions of the truncated theory can be uplifted to solutions 
of the untruncated one, separates formally both issues: that of 
the consistency of the truncation and that of the correctness of the 
uplifting.

Another, less restrictive, concept of
consistent truncation can be found in the literature. A definition
is proposed in \cite{pope} that just proceeds through the e.o.m.:
one starts with a Lagrangian density and introduces some ansatz
for the reduction of the fields, which is then plugged into the
e.o.m.. If the original e.o.m. are compatible with such an ansatz,
the reduced e.o.m. will be considered as a consistent truncation
of the former ones. As a matter of
definition, nothing is wrong in using one meaning or another for
the concept of consistent truncation. We think, however, that the
fulfilment of the commutativity in the diagram above is such a
relevant property that deserves a name by its own, and we find
that {\sl consistent truncation} is the best suited for it, in 
agreement with the approach of many papers in this field. This
is the definition used throughout this paper.

\avall

Sections
\ref{ftt} and \ref{redg}, discuss the Kaluza-Klein
dimensional reduction under a set of independent Killing vector
fields. The main result, the {\sl tracelessness} condition, is
obtained in subsection \ref{e-l-e}, where its sufficiency and
necessity is clarified. The uplifting of solutions is
considered in subsection \ref{upl}, and in section \ref{redg} the
reduction of the gauge group is fully analysed.

Section \ref{fr} considers second-type truncations, which
proceed via the introduction of constraints. In section \ref{appl}
we show with an illustrative example why consistency is difficult
to achieve and also how, despite this lack of consistency, correct
upliftings can possibly be performed in special cases.

\section{First-type truncations: dimensional reduction}
\label{ftt}

\subsection{Killing Symmetries} \label{sec:kill}

Consider a ($d+n$)-dimensional space-time manifold $\mathfrak M$
with a Lorentzian metric tensor $ {\bf
g}_{AB}\,\,(A,B=1,2,\ldots,d+n)$ invariant under a $n$-parameter
group of isometries such that produce $n$-dimensional, space-like,
invariant surfaces (surfaces of homogeneity) that foliate
$\mathfrak M$. This group is generated by $n$
independent space-like vector fields ${\bf K}_{a}$, which span a
Lie algebra:
\begin{equation}
    [{\bf K}_{a},{\bf K}_{b}] = C^{c}_{ab}{\bf K}_{c} \,.
    \label{1.com}
\end{equation}
Every surface of the foliation supports a realization of the Lie
algebra and we think of it as constituting a copy of a
$n$-dimensional sub-manifold $\mathfrak N$. We select local
coordinates in $\mathfrak M$ such that $y^{\alpha}$ are
coordinates along the surfaces of the foliation and $x^\mu$ are
transverse -and will survive the truncation. This means that our
Killing vectors now take the general form
(${\partial}_{\alpha}\equiv\partial/\partial y^{\alpha}$)
\begin{equation}
    {\bf K}_a = K_a\!^{\alpha}(x,y) {}  {\partial}_{\alpha}\,,
\quad \vert a\vert = \vert{\alpha}\vert = n \,.
\end{equation}
In principle we allow an $x$-dependence in $K_a\!^{\alpha}(x,y)$.
We express the isometry property as the invariance of the metric
under the action of the Lie derivative with respect to the vector
field ${\bf K}_a$, ${\mathfrak L}_{{\bf K}_a}({\bf g}) =0$. Other
fields, vectors, tensors, densities, $p$-forms, etc., will be
included in general in our description. The Killing conditions
will be said to be realized on them when they give zero under the
action of the Lie derivative ${\mathfrak L}_{{\bf K}_a}$.

Let us introduce a set of $N$ independent, left-invariant (under the Lie
algebra) vector fields ${\bf Y}_a = Y_a^{\alpha}(x,y) {\partial}_{\alpha}$,
which satisfy
\begin{equation}
    {\mathfrak L}_{{\bf K}_a}{\bf Y}_b =
    \left[{\bf K}_a,{\bf Y}_b\right] = 0\,,
        \label{invarbasis}
\end{equation}
and can be taken to be tangent to the surfaces of the foliation.
With a suitable choice of these vector fields they can be shown to
span the Lie algebra
\begin{equation}
    [{\bf Y}_{a},{\bf Y}_{b}] = - C^{c}_{ab}{\bf Y}_{c}\,.
        \label{commbasis}
\end{equation}

On every surface of the foliation, we define a basis of
one-forms, $ {\omega}^{a} = \omega^a_{\alpha}(x,y) {\bf d}
y^{\alpha}$, dual to the left-invariant vectors: $
{\omega}^{a}\cdot{\bf Y}_{b} = \delta^{a}_{b}$.

Using the one-forms, $ {\omega}^{a}$,
the Lie algebra property \bref{1.com} becomes
\begin{equation}
    {\bf d}_{(y)} {\omega}^{a}
    = \frac{1}{2} C^{a}_{bc}\,
        {\omega}^{b}\! \wedge  {\omega}^{c}\,,
            \label{curlforms}
\end{equation}
being ${\bf d}_{(y)}$ the differentiation with respect to the $y$ variables.

\vspace{6mm}
\noindent
{\it Remark:}
Note that
$$({\mathfrak L}_{{\bf K}_a} {\omega}^{b})\cdot {\bf
Y}_{c} = {\bf K}_a({\omega}^{b}\cdot {\bf Y}_{c})
- {\omega}^{b}\cdot({\mathfrak L}_{{\bf K}_a}{\bf Y}_{c}) = 0
$$
implies
\begin{equation}
\label{alpha}
({\mathfrak L}_{{\bf K}_a} {\omega}^{b}) =
\alpha^b_{a\mu}{\bf d}x^\mu \,,
\end{equation}
for some functions $\alpha^b_{a\mu}$.

\vspace{6mm}

The metric can be written in a more convenient way, using the
mixed basis $\{{\bf d}x^\mu, {\omega}^{a}\}$, partially holonomic,
partially anholonomic, as
\begin{equation}
    {\bf g} =  g_{\mu\nu} {\bf d}x^\mu {\bf d}x^\nu
        + g_{ab}\left(A^a_\mu {\bf d} x^\mu  +  {\omega}^{a}\right)
            \left(A^b_\nu {\bf d} x^\nu +  {\omega}^{b}\right)\,,
                \label{metric}
\end{equation}
where, in principle, all coordinate dependences are allowed,
$g_{\mu\nu}(x,y),\ g_{ab}(x,y)$ and $ A^a_\mu(x,y)$. Notice that
we keep {\it all} components of the metric, for we have $n(n+1)/2$
components for $g_{ab}$, $d(d+1)/2$ for $g_{\mu\nu}$ and $n d$ for
$A^a_\mu$, which sum up to $(d+n)(d+n+1)/2$ independent
components. In this sense, \bref{metric} is just a convenient way
to express the components of the metric, but it does not entail
any reductionist ansatz.

One can indeed explore the conditions on the metric
elements \bref{metric} if we demand them to be
invariant under the action of the Lie derivative ${\mathfrak L}_{{\bf
K}_a}$.
\begin{eqnarray}
{\mathfrak L}_{{\bf K}_a} (g_{bc}) = 0& \Longleftrightarrow& g_{bc} = g_{bc}(x)\,,
        \label{killmetric.a} \\
{\mathfrak L}_{{\bf K}_a}(g_{\mu\nu}{\bf d} x^\mu{\bf d} x^\nu) =
0& \Longleftrightarrow & g_{\mu\nu}=
    g_{\mu\nu}(x) \,,
        \label{killmetric.b} \\
    {\mathfrak L}_{{\bf K}_a}(A^b_\nu {\bf d} x^\nu  + {\omega}^b) = 0
  &  \Longleftrightarrow &
    (\partial_\mu{\bf K}_a\!^{\alpha}) {\partial}_{\alpha}  =
         -{\bf K}_a(A^b_\mu){\bf Y}_b\,.
    \label{killmetric.c}
\end{eqnarray}
The latter relation, \bref{killmetric.c}, links the possible
$x$-dependence of the Killing vectors with the $y$-dependence
of $A^a_\mu$. {F}or the application of the reduction procedure
which is implemented below we shall require this
$y$-dependence in $A^a_\mu$ to cancel out. Therefore we are led to
make the simplification of considering $\partial_\mu{\bf
K}_a\!^{\alpha} =0$, which entails
\begin{equation}
    {\bf K}_a = K_a\!^{\alpha}(y)  {\partial}_{\alpha}\,,
\quad
{\rm and}
\quad
    A^a_\mu = A^a_\mu(x) \,.
\end{equation}
In this case one can also choose the left-invariant vectors ${\bf
Y}_a$ and the one-forms $ {\omega}^a$ to be $x$-independent. Then $
{\mathfrak L}_{{\bf K}_a} {\omega}^{b}=0 \,, $ which means that the
functions $\alpha^b_{a\mu}$ defined in \bref{alpha} vanish. As a
consequence, the metric \bref{metric} is written as
\begin{equation}
{\bf g} =  g_{\mu\nu}(x) {\bf d}x^\mu {\bf d}x^\nu +
g_{ab}(x)\left(A^a_\mu(x) {\bf d}x^\mu  +  {\omega}^{a}\right)
\left(A^b_\nu(x) {\bf d} x^\nu +  {\omega}^{b}\right)\,,
\label{xmetric}
\end{equation}
with
\begin{equation}
     {d\omega}^a
    = {\frac{1}{2}}C^a_{bc}\, {\omega}^b\!\wedge {\omega}^c\,,
\label{domega}
\end{equation}
and the Killing conditions already built in.

\vspace{6mm} \noindent {\it Remark:}  The determinant
of \bref{xmetric}, $\det{g}$, factorises as
\begin{equation}
\det{g} = (\det{g_{\mu\nu}})(\det{g_{ab}}) \vert\omega\vert^2 \,,
\label{factorg}
\end{equation}
being $\vert\omega\vert$ the determinant of $\omega^a_{\alpha}$ .

\vspace{6mm}

Any form can be expressed in terms of the mixed basis, for instance
\begin{equation}
 {\Omega}^{(1)} = \Omega_\mu(x) {\bf d}x^\mu +
\Omega_a(x) {\omega}^{a}\,, \label{oneform}
\end{equation}
\begin{equation}
 {\Omega}^{(2)} = \frac{1}{2} \Omega_{\mu\nu}(x) {\bf d}x^\mu \wedge {\bf
d}x^\nu + \frac{1}{2} \Omega_{ab}(x) {\omega}^{a}\wedge {\omega}^{b} +
\Omega_{\mu a}(x) {\bf d}x^\mu\wedge {\omega}^{a}\,.
\label{twoform}
\end{equation}
In the previous expressions the Killing conditions are built in
automatically because of the single $x$-dependences of the
components. Notice that if the one-form $ {\Omega}^{(1)}$ in
(\ref{oneform}) is interpreted as a Maxwell potential, its field
strength $F^{(2)} = d {\Omega}^{(1)}$ will also satisfy the
Killing conditions, {\it i.e.} after using (\ref{domega}) it will
be cast like (\ref{twoform}). This is a simple consequence of
using (\ref{domega}) because the operation of the exterior
differentiation commutes with that of the Lie derivative. Before
concluding this section we shall prove a useful result.

\vspace{6mm} \noindent {\bf Proposition:}
\begin{equation}
\partial_{\alpha} (\vert\omega\vert Y^{\alpha}_a) = C^b_{ab}\,\vert\omega\vert
\,.
\label{theniceformula}
\end{equation}

\begin{pf}
Take (\ref{domega}),
$$
\partial_{\alpha} \omega^a_{\beta} - \partial_{\beta} \omega^a_{\alpha} =
C^a_{bc}\,\omega^b_{\alpha} \omega^c_{\beta} \,,
$$
saturate it with the inverse matrix to
$\omega^a_{\alpha}$, $Y^{\beta}_d$, and take the trace. It follows
that
$$
{\frac{1}{\vert\omega\vert}} \partial_{\alpha}(\vert\omega\vert)
+ (\partial_{\beta}
Y^{\beta}_a )\omega^a_{\alpha} = C^b_{ab}\,\omega^a_{\alpha} \,.
$$
Saturating again the previous expression with $Y^{\alpha}_d$, we arrive at
\bref{theniceformula}.
\end{pf}

\subsection{Consistent truncations in dimensional reduction}
\label{ctr}

Let us first consider a general variational principle without
implementing any Killing condition on the fields. We are free,
however, to work in the mixed basis
introduced before. Let $\Phi$ symbolise a general component of any
generic field present in our formalism expressed in this basis.
For instance, $\Phi$ may represent $g_{\mu\nu}$, or $A^a_\mu$,
etc. The Lagrangian density will be expressed in terms of the
fields and derivatives. Using the mixed basis for the derivatives\footnote{
We assume a Lagrangian density with up to second
derivatives, but the results trivially generalise to any number of
higher derivatives.},
\begin{equation}
{\mathcal L} = {\mathcal L}\left(\Phi, \Phi_{\mu},\Phi_{\mu\nu},{\bf Y}_b(\Phi),
{\bf Y}_a{\bf Y}_b(\Phi)\right)\,. \label{anhol-var}
\end{equation}
It is worth noticing that the forms $ {\omega}^{a}$ and the
vector fields ${\bf Y}_b$ are not considered field variables in
the Lagrangian; they are just part of the mixed basis used to
express the components of the fields and will never be affected by
the variations defined on the fields in order to formulate
symmetries. The variation of the action,
$$
S = \int d^d x\, d^n y\, {\mathcal L}\,,
$$
produces the Euler-Lagrange equations of motion in the usual way.
Since we are interested in expressing them in the mixed
basis, it is convenient, for reasons to be explained below, to
define $\tilde{\mathcal L}$ by
\begin{equation}
{\mathcal L} =: \vert\omega\vert \tilde{\mathcal L} \,. \label{factor}
\end{equation}

\avall

At the level of the
variational principle, we implement the truncation procedure by
defining the reduced Lagrangian
\[
    {\mathcal L}_R(\Phi, \partial_\mu\Phi,\partial_{\mu\nu}\Phi )
    := \tilde{\mathcal L}(\Phi,
    \partial_\mu\Phi,\partial_{\mu\nu}\Phi,
        \,{\bf Y}_a\Phi =0,\, {\bf Y}_a{\bf Y}_b\Phi = 0)\,.
\]
Note that ${\mathcal L}_R$ is defined as a reduction of
$\tilde{\mathcal L}$, see (\ref{factor}), and {\it not} of
${\mathcal L}$. This is {\it necessary} because in the reduced
Lagrangian the dependences on the $y$ coordinates, located in
$\vert\omega\vert$, must disappear. This can also be seen from the
perspective of directly reducing the action: integrating out the
$y$ coordinates geometrically implies the need for a scalar
density defined in each surface of the foliation;
$\vert\omega\vert$ provides for such a scalar density\footnote{
Notice however that the specialisation on the foliation' surfaces
of the original metric is $g_{ab}\, {\omega}^{a} {\omega}^{b}$,
which shows that the volume of each surface factorises as
$\sqrt{\vert g_{ab}\vert}$ times $\int d^n\!y\vert\omega\vert$.}.

\subsection{Euler-Lagrange equations}
\label{e-l-e}

Once we have introduced the concept of a consistent truncation
in the case of dimensional reductions,
let's inspect it from the point of view of the Euler-Lagrange
equations.
We write the Euler-Lagrange equations using the variables
displayed in (\ref{anhol-var}). Recalling (\ref{factor}), the
variation $\delta{\mathcal L}$ in terms of $\delta\tilde{\mathcal L}$
is\footnote{ The following notation should be understood
hereafter $\Phi_{\mu \nu \ldots}:=
\partial_{\mu\nu \ldots}\Phi$\,.}
\begin{eqnarray}
\delta{\mathcal L} & =&  \vert\omega\vert \delta{\tilde{\mathcal L}} \\& =&
    \vert\omega\vert\left(\frac{\partial{\tilde{\mathcal L}}
    }{\partial \Phi}\delta
    \Phi
        + {\partial{\tilde{\mathcal L}}\over\partial{\Phi_{\mu}}}\delta
        {\Phi_{\mu}}+ \frac{1}{2}{\partial{\tilde{\mathcal L}}\over
        \partial{\Phi_{\mu\nu}}}\delta
        {\Phi_{\mu\nu}}
        ~   + {\partial{\tilde{\mathcal L}}\over\partial{\bf Y}_a\Phi}
            \delta ({\bf Y}_a\Phi)
    + {\frac{1}{2}}
         {\partial{\tilde{\mathcal L}}\over\partial{\bf Y}_a{\bf Y}_b\Phi}
            \delta ({\bf Y}_a{\bf Y}_b\Phi) \right)\,.
\nonumber \label{ltilde}
\end{eqnarray}
 Integration by parts allows us to isolate all pieces with
 $\delta\Phi$ and produces the standard
Euler-Lagrange derivatives. But, in this case, terms involving the
anholonomic basis need some special care. Consider for
instance the term
\begin{eqnarray}
\vert\omega\vert{\partial{\tilde{\mathcal L}}\over\partial{\bf Y}_a\Phi} \delta
({\bf Y}_a\Phi) &=& {\rm divergence} -
\partial_{\alpha}(\vert\omega\vert Y_a^{\alpha} {\partial{\tilde{\mathcal
L}}\over\partial{\bf Y}_a\Phi})
 \delta \Phi\nonumber\\
&=&{\rm divergence}-\vert\omega\vert(C^b_{ab}+{\bf Y}_a){\partial{\tilde{
L}}\over\partial{\bf Y}_a\Phi}
            \delta \Phi \,,
\end{eqnarray}
where we have explicitly made use of the result in
(\ref{theniceformula}).
Applying the same procedure to the rest of the terms in
\bref{ltilde} we arrive at our main result on the
Euler-Lagrange derivatives.

\vspace{6mm}

\noindent {\bf Theorem:} {\sl The Euler-Lagrange equations,
formulated when the field components are expressed in terms of
the mixed basis, take the form}
\begin{eqnarray}
 {\delta{\mathcal L}\over\delta \Phi} & = & \vert\omega\vert
 \left({\partial\tilde{\mathcal L}\over\partial \Phi}
 -\partial_{\mu}{\partial\tilde{\mathcal L}\over\partial\Phi_\mu}+\frac{1}{2}
\partial_{\mu\nu}{\partial\tilde{\mathcal
L}\over\partial\Phi_{\mu\nu}}  \right.\nonumber\\
         & ~ &  -({\bf Y}_a + C_{ac}^c)
               ({\partial\tilde{\mathcal L}\over\partial{\bf
               Y}_a\Phi})
      + \frac{1}{2} \left.
         ({\bf Y}_b+C_{bd}^d)({\bf Y}_a + C_{ac}^c)
          ({\partial\tilde{\mathcal L}\over\partial{\bf Y}_a{\bf Y}_b \Phi})
         \right)\,.
    \label{e-l}
\end{eqnarray}

\vspace{6mm}

In the previous expression the generalisation to higher order
derivatives is straightforward. {F}rom \bref{e-l} it is
immediate the distinction between reducing at the level of the
variational principle versus reducing at the level of the
e.o.m. To manifest this difference sharply we restrict
\bref{e-l} by setting the $y$-derivatives of the fields to zero. In the
following expression this is indicated by the subscript $R$ in
$(X)_{{}_R}$, that stands for $(X)_{_{(\partial_{\alpha} \Phi =
\partial_{{\alpha}{\beta}}\Phi = 0)}}$
\begin{equation}
    \left(\frac{\delta{\mathcal L}}{\delta \Phi} \right)
       _{\!_{\! R}}
     =  \vert\omega\vert\bigg\{\frac{\delta{\mathcal L}_{R}}{\delta \Phi}
        \ - \ C_{ac}^c
          \left({\partial\tilde{\mathcal L}\over\partial{\bf Y}_a\Phi}\right)
            _{\!_{\! R}}
         \ + \ \frac{1}{2}  C_{ac}^c C_{bd}^d
        \left(\frac{\partial\tilde{\mathcal L}}{
            \partial{\bf Y}_a{\bf Y}_b\Phi} \right)
            _{\!_{\! R}}
    \bigg\}\,.
    \label{noncom}
\end{equation}
Equation (\ref{noncom}) displays explicitly the non-commutativity
between the two procedures: reduction of the Lagrangian through
the Killing conditions or reduction of the equations of motion. It
also provides us with
conditions for consistent truncations. To be more
specific,

\vspace{6mm}

\noindent {\bf Theorem (sufficient condition):}
\begin{equation} C_{ac}^c =0\,,\, \forall a \quad
\Longrightarrow \quad \bigg({\delta{\mathcal L}\over\delta \Phi}
\bigg)_{\!_{\! R}}
        = \ \vert\omega\vert\bigg({\delta{\mathcal L}_{R}\over\delta \Phi}\bigg)\,.
\label{statement1}
       \end{equation}
\vspace{4mm}

\noindent {\bf Theorem (necessary and sufficient condition):}
\begin{equation}
\bigg({\delta{\mathcal L}\over\delta \Phi} \bigg)_{\!_{\! R}}
        = \ \vert\omega\vert\bigg({\delta{\mathcal L}_{R}\over\delta \Phi}\bigg)
\quad
\Longleftrightarrow\quad
        \ \ C_{ac}^c
          \left({\partial\tilde{\mathcal L}\over\partial{\bf Y}_a\Phi}\right)
            _{\!_{\! R}}
         \ = \ \frac{1}{2}  C_{ac}^c C_{bd}^d
        \left(\frac{\partial\tilde{\mathcal L}}{
            \partial{\bf Y}_a{\bf Y}_b\Phi} \right)
            _{\!_{\! R}}\,.
\label{statement2}
       \end{equation}
A necessary condition of the type
\begin{equation}
\bigg({\delta{\mathcal L}\over\delta \Phi} \bigg)_{\!_{\! R}}
        = \ \vert\omega\vert\bigg({\delta{\mathcal L}_{R}\over\delta \Phi}\bigg)
\quad
\Longrightarrow\quad
        \ \ C_{ac}^c= 0 \quad  \forall \ a\,,
       \end{equation}
may hold only in specific cases according to the content of the terms
\begin{equation}
          \left({\partial\tilde{\mathcal L}\over\partial{\bf Y}_a\Phi}\right)
            _{\!_{\! R}}
         \,,  \
        \left(\frac{\partial\tilde{\mathcal L}}{
            \partial{\bf Y}_a{\bf Y}_b\Phi} \right)
            _{\!_{\! R}}\,.
       \end{equation}
\vspace{6mm}

The tracelessness condition, $C_{ac}^c =0\,,\, \forall a$, is an
invariant statement, independent of the basis we take for the Lie
algebra, and is equivalent to the statement that the adjoint
representation of the group is unimodular. Abelian Lie algebras
and semi-simple Lie algebras are immediate examples that fulfil
this condition and therefore provide the ground for consistent
truncations. Compact Lie algebras belong also to this class, since
their structure constants can be taken completely antisymmetric
\cite{O'Raifeartaigh:vq}. {F}rom now onwards we shall assume
that the tracelessness condition is satisfied.

\vspace{4mm}

We finish this section with a comment concerning the definition of
${\mathcal L}_R$ as a reduction of $\tilde{\mathcal L}$, see
\bref{factor}, and not of ${\mathcal L}$. We have already argued
that the extraction of a factor $\vert\omega\vert$ was {\sl
necessary}, now we shall show that it is also something {\sl
possible} to be done. Since the Lagrangian is a scalar density,
for the terms where the ``densitisation" is provided by
$\sqrt{\vert g\vert}$ it is clear from (\ref{factorg}) that we can
factorise a term $\vert\omega\vert$. Other terms in the Lagrangian
---sometimes called topological, like the Chern-Simons terms---
are densities because they originate in the action $S$ as an
integration of a form of maximum rank. In this case, since all our
forms satisfy the Killing conditions, they are expressed as in
(\ref{oneform}) or (\ref{twoform}) or generalisations. When we
produce the exterior product of some of these forms such as to get
a form of maximum rank, it is clear that the determinant
$\vert\omega\vert$ will always factorise, thus providing for the
mechanism at work in (\ref{factor}). We give in appendix
\ref{proof} the general proof that the only $y$-dependences in the
Lagrangian ${\mathcal L}$, when the Killing conditions are
operating on the fields, are contained in the determinant
$\vert\omega\vert$.

\subsection{Uplifting of solutions}
\label{upl}

Let us now connect the issue of a consistent truncation in this
case of dimensional reduction with that of the uplifting of
solutions from the lower dimensional theory to the higher
dimensional one.

\vspace{6mm}

It is easy to convince oneself that, using the
notation of the previous section, the following proposition holds:

\noindent {\bf Proposition:}
If a field configuration in $d$
dimensions,
\begin{equation}
\Phi(x) \equiv g_{\mu\nu}(x),\ g_{ab}(x),\ A^a_\mu(x),\
\omega_\mu(x),\ \omega^a(x),\ \omega_{\mu\nu}(x),\ \omega_{\mu
a}(x),\ \omega_{a b}(x)\,, \quad
{\rm etc}.\,,
\label{red-sol}
\end{equation}
is a solution of the reduced equations
$$\left(\frac{\delta{\mathcal L}}{\delta \Phi} \right)
       _{\!_{\! R}} =0\,,$$
then the associated field configuration in $d+n$ dimensions,
\bref{xmetric},\bref{oneform},\bref{twoform}, etc., is a solution
of
$$\left(\frac{\delta{\mathcal L}}{\delta \Phi} \right)=0\,.$$

\avall

Then, recalling \bref{statement1} one arrives at the
following

\noindent {\bf Proposition:} \emph{If the conditions
$C^a_{ab}=0$ for a consistent truncation hold}, we can
assert that if \bref{red-sol} is a solution of the Euler-Lagrange
equations for ${\mathcal L}_{\!R}$, then
\bref{xmetric},\bref{oneform},\bref{twoform}, etc., is a solution
of the Euler-Lagrange equations for ${\mathcal L}$. In other
words, \emph{there is a correct uplifting from the solutions of
the $d$-dimensional theory to solutions of the $(d+n)$-dimensional
theory}.

\subsection{Fermionic sector}

The analysis in section \ref{e-l-e} only deals with bosonic
fields, which is a sector of a Lagrangian that, as we have argued
in the Introduction, stands for a consistent truncation of it. In
view of the potential interest of reducing dimensions in
supersymmetric theories we sketch how fermions are to be
incorporated so that the tracelessness condition is still a
guarantee for the consistency of the truncation.

In this case
we need to define a vielbein in order to couple the space-time
indices with the Lorentzian indices in tangent space. Let us
introduce some notation for the indices. {F}rom now on
$\hat\mu = (\mu, {\alpha})$ denotes the ``curved" indices for the $x$ and $y$
coordinates respectively, whilst $\hat I = (I, A)$ its corresponding
``flat" indices.

A vielbein $e^{\hat I}_{\hat \mu}$ associated with the metric
(\ref{xmetric}) is easily constructed. First, the
$e^I_{\hat\mu}(x)$ are chosen to satisfy
\begin{equation}
e^I_\mu(x)e^J_\nu(x) \eta_{{}_{IJ}} = g_{\mu\nu}(x)  \quad
{\rm and} \quad
e^I_{\alpha}=0\,.
\end{equation}
Next we define $e^A_\mu := b^A_a(x) A^a_\mu(x)$ and $e^A_{\alpha}
:= b^A_b(x) \omega^b_{\alpha}(y)$, where the $b^A_a(x)$ are chosen
such that
\begin{equation}
b^A_a(x)b^B_b(x) \delta_{AB} = g_{ab}(x) \,.
\end{equation}
As a consequence of this construction, the 1-forms $e^{\hat
I}$ of this vielbein already satisfy the Killing conditions,
\begin{equation}
      {\mathfrak L}_{{\bf K}_a}e^{\hat I}_{\hat \mu} = 0   \,.
\end{equation}
Then, after introducing the action of the Lie derivative on the
spinors by means of the Weyl rule, which transforms the spinors as
scalars, we can conclude that, in order to be ready for the
truncation procedure, our spinors must only depend on the
$x$-coordinates. With this implementation of the Killing
conditions on the spinors, the truncation procedure discussed
above extends its consistency to theories with fermions.

\section{First-type truncations: reduction of the Gauge Group}
\label{redg}

We continue with the case of a dimensional reduction satisfying
the conditions for a consistent truncation. The original theory
has a subset of solutions that can be understood as uplifted
from the lower dimensional one. The gauge subgroup that survives
the truncation is given by the elements of the gauge group that
work internally on this subset, that is, that map solutions within
this subset to solutions that are still in it. In the language of
Killing conditions, these are the elements of the original gauge
group that map configurations satisfying the Killing conditions to
configurations that still satisfy them.
\subsection{Abelian Maxwell gauge group}

Suppose that $ {\Omega}^{(1)}$ supports the action of a $U(1)$
gauge group, expressed in the mixed basis as
$$
\delta \Omega_\mu = \partial_\mu \Lambda,\quad\delta \Omega_a =
{\bf Y}_a \Lambda \,.
$$
The gauge group will be reduced by requiring the transformed
objects to satisfy the Killing conditions. In this case, the
requirement that $(\delta\Omega_\mu)$ and $(\delta\Omega_a)$
depend only on the $x$ variables sets $\Lambda(x,y)$ to be
$\Lambda(x,y) = \Lambda(x) + \tilde \Lambda(y)$ with $\Lambda (x)$
arbitrary and $\tilde \Lambda(y)$ such that ${\bf Y}_a \tilde
\Lambda = b_a$ with $b_a$ constant. Using the Lie algebra property
\bref{commbasis} one shows that $b_a$ is restricted to satisfy
$b_a C^a_{bc} =0$. Therefore we see that the reduction of the
Maxwell gauge group is quite trivial: we just obtain the $U(1)$
gauge group for the reduced one-form $\Omega_\mu {\bf d}x^\mu$,
plus a rigid Abelian group of symmetries acting only on the
scalars, $\delta \Omega_a = b_a$, with the condition that $b_a
C^a_{bc} =0$\footnote{For semi-simple Lie algebras of Killing
vectors, this rigid group is void, because $b_a C^a_{bc} =0$
implies in such case that $b_a =0$.}.
\subsection{Diffeomorphism-induced gauge group}

Since we are considering the presence of a metric tensor, we
expect the gauge group in the original theory to include the
diffeomorphisms. Again, in the process of truncation we
automatically produce a partial fixation of the gauge, because
the transformed objects must still satisfy the
original Killing conditions. Take for instance the one-form in
(\ref{oneform}),
\begin{equation}
 {\Omega}^{(1)} = \Omega_\mu(x) {\bf d}x^\mu + \Omega_a(x)
 {\omega}^{a}\,,
\end{equation}
then the reduced diffeomorphisms will be those that produce
variations $\delta  {\Omega}^{(1)}$ such that, {\sl when expressed
in the same mixed basis ${\bf d}x^\mu $ and $ {\omega}^{a}$}, its
components exhibit the coordinate dependences
\begin{equation}
\delta {\Omega}^{(1)} = (\delta\Omega_\mu)(x)\, {\bf d}x^\mu +
(\delta\Omega_a)(x)\,  {\omega}^{a}\,.
\label{delomega}\end{equation} If the --most general--
diffeomorphism variation is generated by the vector field
$$
\vec{v}= \epsilon^\mu(x,y)\partial_\mu + \rho^a(x,y){\bf Y}_a
\equiv \vec{\epsilon} + \vec\rho\,,
$$
(\ref{delomega}) can be alternatively written as
\begin{eqnarray}
\delta {\Omega}^{(1)} &=& {\mathfrak
L}_{\vec{v}} {\Omega}^{(1)} = (\vec{v}\,\Omega_\mu) {\bf
d}x^\mu + \Omega_\mu{\mathfrak L}_{\vec{v}}{\bf d}x^\mu +
(\vec{v}\,\Omega_a)  {\omega}^{a}+\Omega_a {\mathfrak
L}_{\vec{v}} {\omega}^{a}
\nonumber \\
&=&(\vec{\epsilon}\,\Omega_\mu) {\bf d}x^\mu +\Omega_\mu{\bf
d}\epsilon^\mu + (\vec{\epsilon}\,\Omega_a)
 {\omega}^{a}+\Omega_a {\mathfrak L}_{\vec{v}} {\omega}^{a} \,.
\end{eqnarray}
The Lie derivative in the last term is
$$
{\mathfrak L}_{\vec{v}} {\omega}^{a} =\partial_\mu\rho^a{\bf
d}x^\mu + ({\bf Y}_b\rho^a + \rho^c C^a_{cb}) {\omega}^{b} \,,
$$
thus we get
\begin{equation}
(\delta\Omega_\mu)=\vec{\epsilon}\,\Omega_\mu +(\Omega_\nu)
\partial_\mu\epsilon^\nu + \Omega_a\partial_\mu\rho^a \,,
\label{delomega-mu}
\end{equation}
and
\begin{equation}
 (\delta\Omega_b)= \vec{\epsilon}\,\Omega_b
 +\Omega_a({\bf Y}_b\rho^a + \rho^c C^a_{cb}) \,.
\label{delomega-b}
\end{equation}
Now, imposing that (\ref{delomega-mu}) and (\ref{delomega-b}) must
depend only on the $x$-coordinates sets the requirements for the
reduced diffeomorphisms to preserve the Killing conditions that
all our objects satisfy: we get that $\epsilon^a$ must depend on
the $x$-coordinates only, and that $\rho^a(x,y)$ must be of the
form $\rho^a(x,y) =\eta^a(x)+\xi^a(y)$. Thus we have obtained the
nice decomposition
\begin{equation} \vec{v}=
\epsilon^\mu(x)\partial_\mu + \eta^a(x){\bf Y}_a + \xi^a(y){\bf
Y}_a \equiv \vec{\epsilon}+\vec{\eta}+\vec{\xi}\,,
\label{nicedecomp}
\end{equation} with arbitrary $\epsilon^\mu(x)$
and $\eta^a(x)$ and with $\xi^a(y)$ satisfying the condition ${\bf
Y}_b\xi^a + \xi^c C^a_{cb} = - B^a_b$ for some constant matrix
$B^a_b$ (the minus sign is set for later convenience).

The decomposition (\ref{nicedecomp}) has been obtained by
considering the reduction of the diffeomorphism algebra acting on
a one-form and asking for the preservation of the Killing
conditions, but it is important to remark that the result is
general and holds true for whatever form, vector o tensor (in
particular the metric tensor) is analysed in view of the reduction
procedure.

\vspace{4mm}

Notice that (\ref{nicedecomp}) can be summarised in two
mandates, as follows,

 \noindent {\bf Proposition:} The reduction
(\ref{nicedecomp}) of the diffeomorphisms algebra has to preserve

\begin{itemize}
\item[{\it{i})}]
The foliation defined by the Lie algebra of the
Killing vector fields.

\item[{\it{ii})}]
The Lie algebra of the left-invariant
vector fields.
\end{itemize}

\vspace{4mm}

The fact that $\epsilon^\mu$ depends only on the
$x$-coordinates guarantees the preservation of the foliation. As
for the second mandate, note the following actions on  ${\bf Y}_a$
\begin{itemize}
\item[{\it{i})}]
${\mathfrak L}_{\vec\epsilon}{\bf Y}_a =0$\,,
\item[{\it{ii})}]
${\mathfrak
L}_{\vec{\eta}}{\bf Y}_a = -\eta^c C^b_{ca}{\bf Y}_b$\,,
\item[{\it{iii})}]
${\mathfrak L}_{\vec{\xi}}{\bf Y}_a = -({\bf Y}_a\xi^b + \xi^c
C^b_{ca}){\bf Y}_b  = B^b_a{\bf Y}_b$\,.
\end{itemize}
The first result deserves no comments, the second describes the
{\sl inner automorphisms} of the Lie algebra of the
left-invariant vector fields, and the third describes the
{\sl outer automorphisms} of that Lie algebra. Indeed the
integrability conditions for $\xi^a$ to satisfy
\begin{equation}{\mathfrak
L}_{\vec{\xi}}{\bf Y}_a =  B^b_a{\bf Y}_b\,, \label{inner}
\end{equation} for some constant matrix $B^b_a$, are
\begin{equation}
    C_{eb}^a B^e_{c} - C_{ec}^a B^e_{b} + C_{bc}^e B^a_{e} = 0\ ,
        \label{cacaca}
\end{equation}
as one can check by reading equation (\ref{inner}) as ${\mathfrak
L}_{{\bf Y}_a}\vec{\xi} =  - B^c_a{\bf Y}_c$ and applying ${\mathfrak
L}_{{\bf Y}^b}$ to both sides.

Equation (\ref{cacaca})  is the definition of Lie algebra
automorphisms. $B^a_b=0$ is our first case. The trivial case
$B^a_b = - \eta^c C^a_{cb}$ for an arbitrary $\eta^c(x)$
corresponds to our second case, of inner automorphisms, which now
become foliation-dependent, {\sl inner} because they are generated
by the Lie algebra itself. The remaining automorphisms, the {\sl
outer} ones, correspond to the third case, that is, matrices
$B^c_a$ satisfying (\ref{cacaca}) but that {\sl are not} of the
form $\lambda^c C^a_{cb}$ for any constants $\lambda^c$.

\avall

These
last automorphisms were already noticed in \cite{Ashtekar:wa} in
the context of the Bianchi homogeneous cosmologies, where were
given there the name of {\sl homogeneity preserving
diffeomorphisms}.

In the remainder of this section
we shall show that to this decomposition (\ref{nicedecomp})
there are associated three corresponding groups of transformations
acting on the reduced theory, namely,
\begin{itemize}
\item[{\it{i})}]
$\epsilon^\mu(x)\partial_\mu \Longleftrightarrow$
diffeomorphism algebra in the reduced theory,
\item[{\it{ii})}]
$\eta^a(x){\bf Y}_a \Longleftrightarrow$ Yang-Mills algebra of
symmetries in the reduced theory,
\item[{\it{iii})}]
$\xi^a(y){\bf Y}_a
\Longleftrightarrow$ residual algebra of rigid symmetries in the
reduced theory.
\end{itemize}

\subsubsection{Diffeomorphisms in the reduced space-time}
Generators of diffeomorphisms of the type
\begin{equation}
 {\epsilon}= \epsilon^\mu(x)\partial_\mu
\label{reddiff}
\end{equation}
produce the standard space-time diffeomorphisms for the reduced
theory. Under these diffeomorphisms $g_{\mu\nu}$ transform as
tensor components, $A^a_\mu$ as vector components, and $g_{ab}$ as
scalars. Also, if for instance one-forms are present, the components
$\Omega_\mu$ transform as vectors (thus defining a one-form in the
reduced theory) and $\Omega_a$ as scalars. And similarly for
higher order forms.
\subsubsection{The Yang-Mills gauge transformations}
Now consider diffeomorphisms of the type
\begin{equation}
\vec{\eta}= {\eta}^a(x){\bf Y}_a \,. \label{gaugetr}
\end{equation}
Recalling (\ref{delomega-mu}), (\ref{delomega-b}) and similar
relations for the transformation of the metric tensor, we obtain,
under (\ref{gaugetr}):
\begin{eqnarray}
\label{ym}
 \delta_{\mathfrak G}[\vec{\eta}\,] g_{\mu\nu} &=& 0\,, \nonumber \\
\delta_{\mathfrak G}[\vec{\eta}\,] g_{ab} &=& \eta^d
(C_{da}^c g_{cb} + C_{db}^c g_{ac}) \,, \nonumber \\
\delta_{\mathfrak G}[\vec{\eta}\,]A^a_\mu &=&
\partial_\mu\eta^a +  A^c_\mu C_{cd}^a\eta^d \,, \nonumber \\
\delta_{\mathfrak G}[\vec{\eta}\,] \Omega_\mu &=&
 \Omega_a \partial_\mu\eta^a \,,\nonumber \\
\delta_{\mathfrak G}[\vec{\eta}\,]\Omega_a  &=& \eta^d C_{da}^c
\Omega_{c} \,.
\end{eqnarray}
It is worth making a few remarks on \bref{ym}:
\begin{itemize}
\item[{\it{i})}] The third equation identifies $A^a_\mu$ as the
gauge bosons for the Yang-Mills theory\footnote{ The emergence of
a Yang-Mills theory by way of the process of dimensional reduction
was already pointed out by DeWitt, who proposed it
\cite{dewitt;63} as an exercise (problem 77) in his 1963 Les
Houches lectures. It is also known \cite{O'Raifeartaigh:ia} that
in the fifties Pauli worked out a 4-d model obtained after a
reduction of a 6-d one on a $S^2$, exhibiting a $SU(2)$ Yang-Mills
theory, but did not publish it.} associated with the Lie algebra
of the Killing vectors. \item[{\it{ii})}] In quotienting out the
group of Killing symmetries used for the dimensional
reduction, this group resurfaces again in a gauged form.
\item[{\it{iii})}] The fields $g_{ab}$ transform under the
adjoint representation of the gauge group for each index, as does
$\Omega_a$ as well. They describe charged objects for the
Yang-Mills field. Notice, though, that $\det g_{ab}$ is uncharged
because $C_{ab}^b=0$ implies $\delta_{\mathfrak G}[\vec{\eta}\,]
(\det g_{ab}) =0$. \item[{\it{iv})}] The transformation of
$\Omega_\mu$ under the Yang-Mills gauge symmetry does not depend
of $\Omega_\mu$ itself, but only of the scalars $\Omega_a$. This
is of course an artifact of the reduction procedure, but it is
quite unusual.
\end{itemize}

The action of \bref{ym} extends to higher order forms, that will
have brought to the reduced theory more scalar fields. Even in the
Abelian case the p-forms will undergo a Yang-Mills transformation
except for configurations for which their associated scalar fields
--byproduct of the consistent truncation-- vanish.

\subsubsection{The residual rigid symmetries:
 Lie algebra outer automorphisms}

Finally, consider the transformations generated by
$$
\vec\xi = \xi^b(y) {{\bf Y}_b}\,,
$$
with $\vec\xi$ satisfying \bref{inner} and \bref{cacaca}. At first
sight, one could wrongly think that their effect should be erased
during the truncation process -that eliminates the
$y$-coordinates- and thus that they would not be present in the
reduced theory. However, they in fact induce rigid symmetries in
the reduced theory, as one can verify by examining
(\ref{delomega-mu}), (\ref{delomega-b}) and similar relations for
the metric tensor and any other fields. Thus we get the action of
the generators of automorphisms $\delta_{\mathfrak A}[B^a_b]$ on
our fields,
\begin{eqnarray}
 \delta_{\mathfrak A}[B] g_{\mu\nu} &=& 0 \,,\nonumber \\
\delta_{\mathfrak A}[B] g_{ab} &=&
-(B_a^c g_{cb} + B_b^c g_{ac}) \,, \nonumber \\
\delta_{\mathfrak A}[B]A^a_\mu &=&
 B^a_bA^b_\mu\,, \nonumber \\
 \delta_{\mathfrak A}[B] \Omega_\mu  &=& 0\,, \nonumber \\
 \delta_{\mathfrak A}[B] \Omega_a &=& - B_a^b \Omega_b \,.
\label{outerautof}
\end{eqnarray}

\vspace{10mm} The full initial diffeomorphism group has been
reduced by the consistent truncation to three subgroups, namely,
the group of diffeomorphisms in the reduced space, a Yang-Mills
gauge group based on the group of Killing symmetries, and the
rigid group of outer automorphisms of its Lie algebra.
Using the notation $\delta_{\mathfrak D}[\epsilon^\mu],
\delta_{\mathfrak G}[\eta^a], \delta_{\mathfrak A}[B^a_b]$ for the
generators of reduced diffeomorphisms, Yang-Mills gauge
transformations and Lie algebra outer automorphisms respectively,
we have the following commutation relations ($\epsilon^\mu$ and
$\eta^a$ are arbitrary functions of $x$; $B^a_b$ are constant
matrices satisfying (\ref{cacaca}) that are not of the form
$\lambda^c C^a_{cb}$),
\begin{eqnarray}
\big[\delta_{\mathfrak G}[\vec{\eta}\,],\, \delta_{\mathfrak
D}[\vec{\epsilon}\,]
  \big] &=& \delta_{\mathfrak G}[\epsilon^\mu\partial_\mu(\eta^a)]\,,
  \nonumber\\
\big[\delta_{\mathfrak A}[B],\, \delta_{\mathfrak D}[\vec{\epsilon}\,]
\big] &=&
  0\,, \nonumber\\
\big[[\delta_{\mathfrak A}[B],\, \delta_{\mathfrak G}[\vec{\eta}\,] \big]
&=& \delta_{\mathfrak G}[B^a_b\eta^b] \label{comm-rel}\,,
\end{eqnarray}
displaying the structure\footnote{$\otimes$ represents a direct product
and $\wedge$ a semi-direct product.}
$$
({\mathfrak D} \otimes {\mathfrak A}) \wedge {\mathfrak G} \,,
$$
for the final reduction of the initial diffeomorphism group.

Note that in the Abelian case, the group of rigid symmetries
(originated from the outer automorphisms) is just $GL(n,R)$, with
$n$ being the dimension of the group. This is the case of
reductions on a $n$-torus. When the Lie group is semi-simple, its
outer automorphisms are the symmetries of its Dynkin diagram.

\section{Second-type truncations: constraints}
\label{fr}

Hitherto we have only dealt with the consequences of first-type
truncations, that is, dimensional reductions, which are guaranteed
to be consistent due to the tracelessness condition. Notice that
dimensional reductions keep unchanged he number of degrees of
freedom per space-time point. However, in most cases in the
literature, dimensional reduction is accompanied by a second-type
truncation, that entails a true elimination of degrees of freedom
through the introduction of constraints\footnote{The constraints
introduced here are not meant to modify the dynamics, as it would
have generally happened should they had been added to the
Lagrangian with the appropriate Lagrange multipliers, but just to
select a subset of solutions of the original dynamics.}. This is a
totally different matter that has to be dealt with proper methods.
But instead of giving general results on second-type truncations,
which goes beyond the scope of this paper, we shall concentrate in
the study of some constraints commonly adopted in the literature.
We ask in this section whether is possible to achieve a further
simplification by reducing some degrees of freedom in the whole
formalism.  {F}or this purpose we should increase the degree of
symmetry in the Lagrangian. {F}or instance, one can examine
whether it is possible for the metric and/or the $p$-forms
to be simultaneously invariant with respect to the vector fields
${\bf K}_a$ and ${\bf Y}_a$. In this case we say that the metric
is bi-invariant.

Let us examine the consequences of requiring our left-invariant
vector fields ${\bf Y}_a$ to be also Killing for the metric
\bref{xmetric} and for the rest of structures, \bref{oneform},
\bref{twoform}, etc \ldots\,.

\begin{itemize}
\item[{\it{i})}]
Imposing first the Killing condition
\begin{equation}
\label{kilY}
{\mathfrak L}_{{\bf Y}_c} {\omega}^{a}= C^a_{cb}\
{\omega}^{b} \,,
\end{equation}
on the metric we obtain
\begin{equation}
{\mathfrak L}_{{\bf Y}_c}{\bf g} = g_{ab}(x)\left( {\mathfrak L}_{{\bf
Y}_c}{\omega}^{a}\right) \left(A^b_\nu(x) {\bf d} x^\nu +
{\omega}^{b}\right) + g_{ab}(x)\left(A^a_\mu(x) {\bf d}x^\mu  +
{\omega}^{a}\right) \left( {\mathfrak L}_{{\bf
Y}_c}{\omega}^{b}\right) = 0\,.
\label{2fase}
\end{equation}
After substituting \bref{kilY} the last expressions amounts to
\begin{equation}
g_{ab}C^a_{cd} + g_{ad}C^a_{cb} = 0\,, \label{2fase-metr-cond}
\end{equation}
and
\begin{equation}
g_{ab}A^a_\mu C^b_{cd}  = 0\,. \label{2fase-metr2-cond}
\end{equation}
In turn, using \bref{2fase-metr-cond}, we can express
\bref{2fase-metr2-cond} as
\begin{equation}
A^a_\mu C^b_{ac}  = 0\,. \label{2fase-metr3-cond}
\end{equation}
Condition \bref{2fase-metr3-cond} can indeed be very restrictive,
for instance for semi-simple Lie algebras, for which the
Cartan-Killing constant metric $h_{ab} = C^c_{ad}C^d_{bc}$ has
determinant $\det{h_{ab}}\neq 0$, it implies the vanishing of the
YM gauge potential.

\item[{\it{ii})}] On the other hand the implementation of the new
Killing conditions on the one-form \bref{oneform} become
\begin{equation}
\Omega_a C^a_{cd} = 0\,, \label{2fase-onef-cond}
\end{equation}
whereas for the two-form \bref{twoform} we have
\begin{equation}
\Omega_{\mu a} C^a_{cd} = 0\, \quad
{\rm and} \quad
\Omega_{ab}C^a_{cd} - \Omega_{ad}C^a_{cb} = 0\,.
\label{2fase-twof-cond}
\end{equation}
\end{itemize}

\vspace{4mm}

Summing up, the requirement of bi-invariance of the metric may set
a strong restriction on the gauge potential, which indeed vanishes
in the case of semi-simple algebras. The conclusion is that such a
requirement, as a way to produce further reductions, is too
stringent and not sustainable in the most interesting cases. A
milder requirement, worth to explore, is still to demand that all
our objects --or some of them-- be bi-invariant, but only
when specialised on any surface of the foliation. In the case of
the metric, its specialisation to the surface labelled by the
coordinates $x$ is $ g_{ab}(x) {\omega}^{a} {\omega}^{b} $, and
the condition for bi-invariance is just \bref{2fase-metr-cond}.
Similarly, the conditions for bi-invariance of the one-forms and
two-forms, specialised to any surface of the foliation, are
\bref{2fase-onef-cond} and \bref{2fase-twof-cond}. Therefore, with
this slight change, the unwanted condition \bref{2fase-metr3-cond}
no longer shows up.

It is interesting to notice that the remaining conditions,
\bref{2fase-metr-cond},\bref{2fase-onef-cond},\bref{2fase-twof-cond},
bring a remarkable simplification to the reduced gauge group.
Particularly, the Yang-Mills gauge group \bref{ym} becomes
\begin{eqnarray}
 \delta_{\mathfrak G}[\vec{\eta}\,] g_{\mu\nu} &=& 0\,, \nonumber \\
\delta_{\mathfrak G}[\vec{\eta}\,] g_{ab} &=& 0 \,, \nonumber \\
\delta_{\mathfrak G}[\vec{\eta}\,]A^a_\mu &=&
\partial_\mu\eta^a +  A^c_\mu C_{cd}^a\eta^d \,, \nonumber \\
\delta_{\mathfrak G}[\vec{\eta}\,] \Omega_\mu &=&
 \Omega_a \partial_\mu\eta^a \,,\nonumber \\
\delta_{\mathfrak G}[\vec{\eta}\,]\Omega_a  &=& 0 \,. \label{ym2}
\end{eqnarray}

\vspace{4mm}

\noindent {\it Remark:}  Eq.~\bref{ym2} implies that the possible
scalars that still remain as independent degrees of freedom are
neutral with respect to the YM interaction, that is, there is no
minimal coupling. Thus, \emph{the requirement of bi-invariance
on the surfaces of the foliation amounts to removing all the
charged scalars from the formalism}.

\vspace{4mm}

Conditions \bref{2fase-metr-cond}, \bref{2fase-onef-cond},
\bref{2fase-twof-cond} are just constraints to be implemented in
the formalism. They amount to the elimination of some --or many--
of the scalars that have appeared in the theory via
dimensional reduction. In the following we shall impose
these constraints, and analyse whether such a procedure is still a
consistent truncation. Note nevertheless that one can choose to
impose this bi-invariance requirement only on the metric field,
and not the other structures (p-forms for instance).

Implementation of constraints may or may not further reduce the
gauge freedom avaible to the field theory. In the former case,
such constraints as known as gauge fixing constraints. The
consequences of their implementation have been studied in
\cite{plugging}. In our case, however, the constraints considered
above do not fix any gauge as the following proposition shows.

\vspace{4mm}

\noindent {\bf Proposition:}
The introduction of the constraints
\bref{2fase-metr-cond}, \bref{2fase-onef-cond},
\bref{2fase-twof-cond}
does not imply further restrictions of the gauge group. This results from
the fact that the transformations
\bref{reddiff}, \bref{ym} and \bref{outerautof} already preserve
the constraints.

\vspace{4mm}

\noindent \begin{pf} Since the constraints are linear and only
involve scalar fields, it is obvious that they are preserved by
the reduced diffeomorphisms \bref{reddiff}. They are also
preserved by the relevant transformations in \bref{ym} because,
when acting on the scalar fields, these transformations are
linear. Finally, concerning the outer automorphisms, notice that,
using \bref{ym}, we can express the constraints as
$$\delta_{\mathfrak
G}[\vec{\eta}_0] g_{ab} =0\,,\quad \delta_{\mathfrak G}[\vec{\eta}_0]
\Omega_a =0\,,\quad \delta_{\mathfrak G}[\vec{\eta}_0] \Omega_{ab} =0\,,$$ for
$\vec{\eta}_0 =$ any constant vector. Then use of the commutation
relations \bref{comm-rel} guarantees that the transformations
\bref{outerautof} preserve the constraints.\end{pf}

\vspace{4mm}

It is worth noticing that there is always a solution for the
system of constraints \bref{2fase-metr-cond}.

\vspace{4mm}

\noindent {\bf Proposition:} The Cartan-Killing metric $h_{ab}$
automatically satisfies \bref{2fase-metr-cond} {\it i.e.}
\begin{equation}
h_{ab}C^a_{cd} + h_{ad}C^a_{cb} = 0\,.\label{h-cond}
\end{equation}

\vspace{4mm}

\noindent \begin{pf} The l.h.s. of \bref{h-cond} is
$$
C^e_{af}C^f_{be}C^a_{cd}+C^e_{af}C^f_{de}C^a_{cb} =
C^f_{be}(C^e_{ac}C^a_{fd}+C^e_{ad}C^a_{cf}) +
C^f_{de}(C^e_{ac}C^a_{fb}+C^e_{ab}C^a_{cf})\,,
$$
where we have used the Jacobi identity. Notice that the first and
fourth terms cancel, as do the second and third as well.\end{pf}

\vspace{4mm}

Constraints
\bref{2fase-metr-cond},\bref{2fase-onef-cond},\bref{2fase-twof-cond},
and the similar ones corresponding to other possible fields
present in the theory, are indeed holonomic constraints, that is,
constraints in configuration space. Now, given a specific theory,
we face two new questions. The first one is as to whether the
e.o.m. are compatible with the constraints. If that is the case
and we use the constraints in the e.o.m., the next point one may
rise is as to whether these reduced e.o.m. are derivable from the
new reduced Lagrangian, which is obtained by plugging the
constraints into the Lagrangian resulting from the first-type
truncation performed before. These questions will be answered
in the next sections.

Notice that this second-type truncation --by implementing
constraints-- has qualitative features that make it very distinct
form the first one. The consistency condition obtained for
first-type truncations was model-independent and only relying on a
property of the Lie algebra of independent Killing vectors,
namely, the tracelessness condition. Contrariwise, in implementing
the new constraints, consistency will be model dependent and it is
most likely that it needs to be checked in a case by case basis.

\subsection{An example: semi-simple Lie algebra}

To be more specific and substantiate the importance of these
second-type truncations, we shall consider in the remainder of
this section an example based on a semi-simple Lie algebra. Let us
then study the fate of the scalars ${g_{ab}}$ under the
bi-invariance constraint \bref{2fase-metr-cond}\footnote{This constraint
has been already presented in \cite{jadczyk}, although without analysing its
dynamical implications.}.

\avall

We start by defining the matrices $D^a_b$ such that one can
express $g_{ab}= D_a^c h_{cb}$, and look for the conditions on
these matrices $\bf D$. {F}rom \bref{2fase-metr-cond} and
\bref{h-cond},
$$
- g_{bf} C^f_{cd} = g_{ad} C^a_{cb} \rightarrow - D_b^e
h_{ef}C^f_{cd}  = D^f_a h_{df} C^a_{cb} \rightarrow D_b^e
h_{df}C^f_{ce}  = D^f_a h_{df} C^a_{cb}\,,
$$
which for a semi-simple lie algebra is equivalent to
$D_b^e C^f_{ce}  = D^f_a  C^a_{cb}$\,, that is
\begin{equation}
[{\bf D},\, {\bf
C}_c ] = 0\,, \label{comm}
\end{equation}
where ${\bf C}_c$ are the matrices of the adjoint representation
$({\bf C}_c)^a_b := C^a_{cb}$\,.
Bearing in mind that a semi-simple lie algebra decomposes as a
direct sum of its simple subalgebras, and that this makes the
adjoint representation to decompose into a direct sum of
irreducible representations (which are the adjoint representations
for the simple subalgebras), Shur's lemma will imply that the
matrix $\bf D$ is a multiple of the identity on every invariant
subspace. The Cartan-Killing metric is also reducible to the
invariant subspaces. The multiplicity for $\bf D$ in each subspace
is arbitrary and, therefore, we end up with {\sl as many scalars
(from the reduction of the metric) as there are simple subalgebras
in the decomposition of our semi-simple subalgebra}. In
particular, when the algebra is simple, the constraint
\bref{2fase-metr-cond} is therefore equivalent to
\begin{equation} g_{ab} = \varphi\, h_{ab} \,,
 \label{2fase-metr-cond2}
\end{equation}
with $\varphi$ being the only remaining scalar from the original
metric. Let us mention that choices of scalars of this type are
common, as ansatzs on the form of the metric. Here we have
presented a rationale for such ansatzs, which is based on
requiring: {\it i}) the Killing symmetry of the metric (and in the
rest of the fields) and {\it ii}) the implementation of the
constraints originated from the bi-invariance of the metric field
on the surfaces of the foliation.

Finally we notice that the rigid symmetry generated by the outer
automorphism \bref{outerautof} gets also notably simplified with
the constraints in the semi-simple case. It only remains a rigid
symmetry acting on the YM potential. On every invariant space
(where $g_{ab}$ satisfies \bref{2fase-metr-cond2}), the r.h.s. of
the transformation $\delta_{\mathfrak A}[B] g_{ab}$, will be $
B_a^c h_{cb} + B_b^c h_{ac}$, that can be shown to vanish by
repeated use of \bref{cacaca}.

\avall

In contrast with the preceding
findings for the metric field, the constraints imposed by the
Killing conditions --with a semi-simple Lie algebra-- on any p-form
set their associated scalars to zero.
The proof is presented in appendix \ref{scalars}.

\section{Application: the Einstein-Hilbert action}
\label{appl}

In this section we analyse, for a specific model, the consequences
of implementing second-type truncations in addition to dimensional
reductions, already undertaken in the preceding section. We
consider the Lie algebra of Killing vectors to be simple. This
means that
\begin{equation}
g_{ab}C^a_{cd} + g_{ad}C^a_{cb} = 0 \quad\Longleftrightarrow \quad
g_{ab}= \varphi h_{ab} \label{simple-case}
\end{equation}
will hold in the sequel.

Even if the attention is restricted to cases of independent
Killing vectors, an analogous analysis \emph{must} be taken into
account in more cumbersome cases such as spherical reductions
\cite{pope}. To the best of our knowledge, this implementation is
not obvious in the present studies.

As a matter of notation, caret quantities will denote coordinate
indices as well as fields of the $d+n$ dimensional theory. The
metric signature convention is $(-++++\cdots)$\,.

The simplest model to study, that will already produce
results applicable to models with more field contents, is Einstein
GR. The $(d+n)$-dimensional Lagrangian is a
pure gravitational one, in terms of the action
\begin{equation}
\label{einstein} S^{(d+n)} = \frac{1}{2\kappa^2} \int
d^dx\,d^ny\,{\vert{-\hat{g}_{\hat\mu\hat\nu}}\vert}^{1/2}
\,\hat{{\mathfrak R}}\,,
\end{equation}
which, after substitution of \bref{xmetric},
\begin{eqnarray}
\label{matrix}
\hat{g}_{\hat\mu \hat\nu}
 = \, \left( \begin{array}{cc}
g_{\mu \nu}\,+\,g_{ab}\, A^a_\mu\, A^b_\nu & g_{ab}\, A^a_\mu\,
\omega^b_\beta \\
g_{ab}\, A^a_\nu\, \omega^b_\alpha &
g_{ab}\, \omega^a_\alpha\, \omega^b_\beta
\end{array}  \right)\,,
\end{eqnarray}
and factorisation of
$\int \vert\omega\vert$, becomes the reduced action \cite{Scherk:1979zr,
chofreund}\footnote{Notice that in \cite{Scherk:1979zr} the $g_{\mu \nu}$
coefficient in \bref{matrix} is multiplied by a power of $\vert\ g_{ab}\vert$
in order to stick the action to the Einstein frame.}
\begin{eqnarray}
S^d& =& \frac{1}{2\kappa^2} \int d^d x\, \vert -
g_{\mu\nu}\vert^{1/2} \vert g_{ab}\vert^{1/2} \Bigg\{ {\mathfrak
R}- \frac{1}{4} F^{\mu\nu a}\, F_{\mu\nu}^b\, g_{ab}\nonumber
+\frac{1}{4} g^{\mu\nu}\, {\mathcal D}_\mu g_{ab}\, {\mathcal
D}_\nu g^{ab}
\\&&
+g^{\mu\nu}\, {\mathcal D}_\mu \ln \sqrt{g_{ab}} \,
{\mathcal D}_\nu \ln \sqrt{g_{ab}}
-\frac{1}{4}\,C^a_{bc}\left[ 2C^b_{a c^\prime} \, g^{c c^\prime} +
C^{a^\prime}_{b^\prime c^\prime}\, g_{aa^\prime}\, g^{bb^\prime}
\,g^{cc^\prime} \right] \Bigg\} \,, \label{ss-formula}
\end{eqnarray}
where $g^{ab}$ and $g^{\mu\nu}$ are the inverse matrices of
$g_{ab}$ and $g_{\mu\nu}$ respectively. The associated
Lagrangian density, $ {\mathcal L}_R$, is a first-type
consistent truncation of $ {\mathcal L}$.

Let us now proceed to the second-type truncation. The
implementation of the constraints \bref{2fase-metr-cond} in
${\mathcal L}_R$ takes the explicit form \bref{simple-case}. This
produces the new Lagrangian (taking $\kappa^2 = 1$)
\begin{equation}
{\mathcal L}_{2R} =
\frac{1}{2}\vert - g_{\mu\nu}\vert^{1/2}\,\varphi^{n/2}\,
 \vert h_{ab}\vert^{1/2} \Bigg\{
{\mathfrak R} - \frac{1}{4} \varphi\, F^{\mu\nu a}\,
F_{\mu\nu}^b\, h_{ab} +\frac{n(n-1)}{4} g^{\mu\nu}\, \partial_\mu
\ln\varphi \,\partial_\nu \ln\varphi +2n\,
\varphi^{-1}\Bigg\}\,. \label{our-formula}
\end{equation}
The task is to examine the possible consistency of the
truncation ${\mathcal L}_R \rightarrow{\mathcal L}_{2R}\,.$

We use the notation $\frac{\delta {\mathcal L}}{\delta X}$ for the
Euler-Lagrange derivatives of any Lagrangian ${\mathcal L}$ with
respect to the field component $X$, and $\left(\frac{\delta {\mathcal
L}}{\delta X}\right)_{\!\varphi}$ to represent the application of
\bref{simple-case} on it. It turns out that
\begin{eqnarray}
&&\left(\frac{\delta {\mathcal L}_R}{\delta
g_{\mu\nu}}\right)_{\!\varphi} = \frac{\delta {\mathcal
L}_{2R}}{\delta g_{\mu\nu}}\,, \nonumber \\&& \label{second-red}
\left(\frac{\delta {\mathcal L}_R}{\delta
A^a_{\mu}}\right)_{\!\varphi} = \frac{\delta {\mathcal
L}_{2R}}{\delta A^a_{\mu}}\,,\\&& \left(\frac{\delta {\mathcal
L}_R}{\delta g_{ab}}\right)_{\!\varphi} =
\frac{1}{n}\left(\frac{\delta {\mathcal L}_{2R}}{\delta
\varphi}\right)h^{ab} -
\frac{1}{2}{\vert{g_{\mu\nu}}\vert}^{\frac{1}{2}}
\varphi^{\frac{n}{2}}{\vert{h_{ab}}\vert}^{\frac{1}{2}}
\left(\frac{1}{4}F^{\mu\nu a}F_{\mu\nu}^{\ b}-
\frac{1}{4n}(F^{\mu\nu c}F_{\mu\nu}^{\ d}\, h_{cd}) h^{ab}
\right)\,. \nonumber
\end{eqnarray}
It is interesting to observe, from \bref{second-red}, that
\begin{equation}
h_{ab}\left(\frac{\delta {\mathcal L}_R}{\delta
g_{ab}}\right)_{\!\varphi} = \left(\frac{\delta {\mathcal
L}_{2R}}{\delta \varphi}\right)\,,
\label{red-eq}
\end{equation}
which holds for any reduction
${\mathcal L}_R\rightarrow {\mathcal L}_{2R}$ that implements
\bref{simple-case}. It can
be proved by just noticing that ${\mathcal L}_{2R}(\varphi,
\dot\varphi,...) := {\mathcal L}_{R}(g_{ab}=\varphi h_{ab},\ \dot
g_{ab}=\dot\varphi h_{ab},\ldots)$\,.

\avall

The last equality in \bref{second-red} reflects a mismatch between
the truncation (that is, the implementation of
\bref{2fase-metr-cond}) of the Euler-Lagrange equations for
${\mathcal L}_R$ and the Euler-Lagrange equations for ${\mathcal
L}_{2R}$. As a consequence, notice that the diagram applicable
to this case, analogous to that sketched in section
\ref{sec:intro}, is \indiag(1,1) \capsa(0,0){{\delta{\mathcal
L}_R\over\delta \Phi}\!=0\ } \capsa(0,1){{\mathcal L}_R}
\capsa(0,1){\flS{{\rm e.o.m.}}} \capsa(0,0){\flE{{\rm Red}.}}
\capsa(0,1){\qquad\quad\flE{{\rm Red}.}} \capsa(1,0){\qquad
\hspace{4.5cm}\ \big(\frac{\delta{\mathcal L}_R}{\delta \Phi}
\big)
       _{\!_{\!{\rm red}}}\!=0\Leftrightarrow\frac{\delta{\mathcal
L}_{\!2R}}{\delta \Phi} +  {\rm new\ \, terms}=0.}
\capsa(1.5,1){{\mathcal L}_{2R}} \capsa(1.5,1){\flS{{\rm e.o.m.}}}
\exdiag
 \noindent
The presence of these new terms is harmless only in the case we
require that the solutions of the e.o.m. for ${\mathcal L}_{2R}$
force them to vanish\footnote{ {F}or the sake of notational simplicity we have
suppressed the space-time indices in the Yang-Mills fields
strengths $F^{a}$.}. That is in our case
\begin{equation}
F^a F^{b}-\frac{1}{n}(F^{c}F^{d}\, h_{cd}) h^{ab} =0\,.
\label{sec-constr}
\end{equation}

\vspace{4mm}

Let us now explain the origin of equation \bref{sec-constr}.

The imposition of the constraints \bref{2fase-metr-cond} on the
solutions of the Lagrangian ${\mathcal L}_R$ triggers a {\sl
stabilisation mechanism} of the type considered in
Dirac-Bergmann's theory of constraints systems
\cite{dirac1,dirac2,bergm1,bergm2,bergm3}: the consistency of the
dynamics generated by ${\mathcal L}_R$ with the constraints
\bref{2fase-metr-cond} may lead to the appearance of new
constraints and/or some new restrictions on the gauge freedom
present in ${\mathcal L}_R$. This latter case does not occur in
our setting, since we have already observed in the previous
section that the constraints \bref{2fase-metr-cond} are preserved
by the transformations \bref{reddiff},\bref{ym},\bref{outerautof}.
The only analysis we need to perform is that of the compatibility
of \bref{2fase-metr-cond} with the dynamics, i.e. the search for
new constraints. For the initial conditions $t=0$ consider the
configuration $g_{ab}(x^i,0)$ and $\dot
g_{ab}(x^i,0)$\footnote{Dot stands for time derivative.}
satisfying \bref{2fase-metr-cond} ($\mu = \{ i, 0 \}$). Then one
should check whether $\ddot g_{ab}(x^i,0)$ still satisfies
\bref{2fase-metr-cond}. Using the equations of motion
$\frac{\delta {\mathcal L}_R}{\delta g_{ab}}=0$ one can isolate
$\ddot g_{ab}(x^i,0)$ in terms of the initial conditions for all
the fields. Then requiring\footnote{ It is more convenient to use
this version of the constraints, equivalent to
\bref{2fase-metr-cond}.}
$$\ddot g^{ac}(x^i,0)C^b_{cd} + \ddot g^{bc}(x^i,0)C^a_{cd} = 0\,,
$$
we obtain
\begin{equation}
F^a F^c C^b_{cd} + F^b F^c C^a_{cd} = 0\,,
\label{constr-on-f}
\end{equation}
valid not only at $t=0$ but, as \bref{2fase-metr-cond}, at any time as well.
We have already argued, using \bref{comm}, that for a simple
Lie algebra, this relation \bref{constr-on-f} is equivalent to
$$
F^a F^b = \alpha h^{ab}\,,
$$
for some function $\alpha$. Then $\alpha$ is determined by saturating
this last relation with $h_{ab}$ and we verify that
\bref{constr-on-f} is nothing but \bref{sec-constr}.

Borrowing the terminology of Dirac-Bergmann's theory of
constrained systems, \bref{2fase-metr-cond} are interpreted as
\emph{primary} constraints and \bref{sec-constr} as
\emph{secondary}, the latter being dynamical consequences of
the stabilisation of the former. In principle the Dirac-Bergmann
algorithm could be continued to get tertiary and higher order
constraints but, concerning a given solution of the ${\mathcal
L}_{2R}$ theory, if this solution satisfies the secondary
constraints {\sl at any time}, it is a priori guaranteed that it
will already satisfy all its dynamical consequences: tertiary
constraints, etc.

\subsection{Uplifting of solutions}
\label{upl2}

Now comes the crucial point in second-type truncations, the issue
of uplifting of solutions. Let us go back to \bref{second-red} (or
the diagram above, from where we borrow the notation), and
consider a solution of the ${\mathcal L}_{\!2R}$ theory that
luckily happens to satisfy the constraints \bref{sec-constr}.
According to \bref{second-red}, this solution will also satisfy
the equations ($\Phi$ represents any field of the ${\mathcal L}_R$
theory)
$$\big(\frac{\delta{\mathcal L}_R}{\delta \Phi} \big)
       _{\!_{\!{\rm red}}}\!=0\,.$$
Now the question remains as to whether this solution can be
uplifted to a solution of
$$\frac{\delta{\mathcal L}_R}{\delta \Phi}
      =0\,.$$
The answer is that this will be possible if the dynamics generated
by ${\mathcal L}_R$, acting on the field configurations of such an
uplifted solution at a given time, preserves the primary
constraints \bref{2fase-metr-cond}\footnote{Thus the vector field
that generates the dynamics is tangent to these constraints, for
the specific field configuration under consideration.} that are
the cause of the truncation. But the condition for the
preservation of the primary constraints is the fulfilling of the
secondary constraints \bref{sec-constr}. Since our solution is
satisfying them by construction, \emph{it is guaranteed that its
uplifting provides us with a solution of the ${\mathcal L}_R$
theory}.

Note the possible existence of a particular case, i.e., when the
Euler-Lagrange equations for ${\mathcal L}_{\!2R}$ directly imply
the constraints \bref{sec-constr}. Only in this very fortunate and
unlikely case the second-type truncation is consistent by its own,
and thus only in this case there is guarantee that every solution
of the ${\mathcal L}_{\!2R}$ theory can be uplifted to a solution
of the ${\mathcal L}_{\!R}$ theory.

\avall

We end up therefore with the following results:
\begin{itemize}
\item[{\it{i})}] A second-type truncation will usually be
inconsistent\footnote{Except for trivial cases as the one
discussed in the introduction.} because of the appearance of residual terms as
depicted in the last diagram (right, bottom side).
These terms are interpreted as secondary constraints.
\end{itemize}

\begin{itemize}
\item[{\it{ii})}] Nevertheless, {\sl a solution of ${\mathcal
L}_{2R}$ can still be uplifted to a solution of ${\mathcal L}_{R}$
if and only if it satisfies the secondary constraints
(\bref{sec-constr} in our example).} We believe that this fact,
that is, the presence of constraints as a condition for the
uplifting under second-type truncations, has not been clearly
stated in the literature.
\end{itemize}
\avall

As a check of consistency, let us address the gauge invariance
properties of the secondary constraints \bref{sec-constr}.
Recalling \bref{ym} one obtains the usual transformation of the YM
field strength
$$
\delta F^a_{\mu\nu}  = \eta^d C^a_{cd} F^c_{\mu\nu}\,.
$$
This, in turn, implies that $F^a F^b$ is gauge invariant \emph{if}
\bref{constr-on-f}
is satisfied, thus guaranteeing that the constraints \bref{sec-constr}
are gauge invariant.

\avall

\begin{figure}
\begin{center}
\epsfig{file=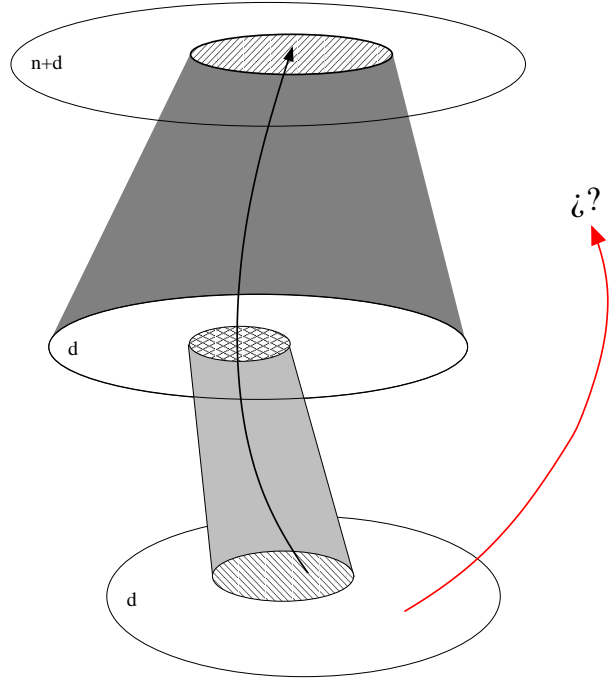,width=8cm,height=9cm}
\end{center}
\caption{Pictorial representation of an entire truncation scheme.
{F}rom top to bottom: the initial truncation corresponds to a
dimensional reduction, $d+n \rightarrow d$, with ${\mathcal
L}\rightarrow{\mathcal L}_{R}$. The upper shadowed area indicates
the solutions of the original theory that fulfil the Killing
conditions. The second-type truncation is performed in a second
step, ${\mathcal L}_{R}\rightarrow{\mathcal L}_{2R}$. It concerns
only with a reduction of the degrees of freedom in the theory,
thus the dimensionality of the space-time is maintained. The total
intermediate region describes the solutions of the ${\mathcal
L}_{R}$ theory, whereas the shadowed area inside corresponds to
the solutions that in addition satisfy some given primary
constraints (in our example, the constraints
\bref{2fase-metr-cond}). The lower region describes the solutions
of the ${\mathcal L}_{2R}$ theory and the shadowed area in it
describes those that can be uplifted to solutions of the
${\mathcal L}_{R}$ theory because they satisfy the secondary
constraints (in our example, the constraints \bref{sec-constr}).}
\label{fig}
\end{figure}

We can summarise the second-type truncation procedure, ${\mathcal
L}_{R}\rightarrow{\mathcal L}_{2R}$, as follows: we introduce some
primary constraints, like \bref{2fase-metr-cond} in our example,
that select a subset of solutions, satisfying the constraints,
among the solutions of the ${\mathcal L}_{R}$ theory. Implementing
directly the constraints in the Lagrangian ${\mathcal L}_{R}$
defines the secondly reduced Lagrangian ${\mathcal L}_{2R}$. When
formulating its e.o.m., there appears a mismatch between the
e.o.m. for ${\mathcal L}_{2R}$ and the implementation of the
constraints into the e.o.m.\! for the Lagrangian ${\mathcal
L}_{R}$. \emph{The cause of this mismatch are the secondary constraints},
like \bref{sec-constr} in our example, that are dynamically
derived from requiring the primary ones to be preserved under the
dynamics, and is reflected in \bref{second-red}. The solutions to
the e.o.m of the ${\mathcal L}_{2R}$ theory that satisfy the
secondary constraints are those that can be uplifted to solutions
of the ${\mathcal L}_{R}$ theory. We sketch these results in
fig.~\ref{fig}. As a conclusion, we remark that, {\sl under
additional conditions (constraints), a suitably prepared
uplifting may be correct even in cases where the reduction
procedure (${\mathcal L}_{R}\rightarrow{\mathcal L}_{2R}$ in our
case) is not a consistent truncation by its own}.

\avall

In the specific example worked out in this section, we have not
proved that such possibility of uplifting is indeed realised. As a
matter of fact, it is not difficult to prepare non-trivial
configurations for a $SU(2)$ gauge potential verifying
\bref{sec-constr}. However, there is no aim in this paper to
produce specific solutions, but rather to portray the standard
picture associated with second-type truncations, and to trace the
origin of possible obstructions to their consistency to the
presence of secondary constraints. The addition of more fields in
the theory, with the subsequent redefinition of the secondary
constraints, may help in bringing more flexibility in order to
ensure that field configurations exist that satisfy the secondary
constraints and the e.o.m. for ${\mathcal L}_{2R}$ altogether.

In fact, it is straightforward to generalise the above
results to theories subjected to the same primary constraints but
with an additional field content. The only not
obvious piece in the analysis is that due to the derivatives
with respect to $g_{ab}$ in \bref{second-red}. We shall cope with
it in the following. Under the same type of constraints, the
e.o.m. of ${\mathcal L}_R$ encode two kinds of structures: one
proportional to the Cartan-Killing metric ${\bf h}$ and the second
involving products of the field strength with two of the Lie
indices $(a,b)$ open. Pulling out terms form the former structure
one can construct in the latter quantities that are YM
gauge-invariant scalar densities with the property of being
orthogonal to $h_{ab}$. Pictorially the set of solutions can be
casted as
\begin{equation}
\label{general}
\left(\frac{\delta {\mathcal L}_R}{\delta g_{ab}}\right)_{\!\varphi} =
s_\parallel^{ab} + s_\perp^{ab}\,,
\end{equation}
with $$s_\parallel^{ab} := \left(\frac{\delta {\mathcal
L}_{2R}}{\delta \varphi}\right) h^{ab} \,,\, \quad s_\perp^{ab}
h_{ab} \equiv 0\,,$$ and $s_\perp^{ab}$,which is an equivalent
way to express the secondary constraints, may contain in
principle the most general structure compatible with \emph{all}
the symmetries of the nature of the problem. It is therefore
evident that, as regards the e.o.m., the truncation
procedure is equivalent to a projection of the e.o.m. of the
initial ${\mathcal L}_R$ theory. This is achieved with the
Cartan-Killing metric acting as a projection-like operator. Thus
in a second-type truncation, here represented by ${\mathcal
L}_{R}\rightarrow{\mathcal L}_{2R}$, the information from these
terms in $s_\perp^{ab}$ is lost. The real problem then appears in
the uplifting procedure: starting with a generic solution of the
${\mathcal L}_{2R}$ theory one can only be confident to reobtain a
solution of the initial ${\mathcal L}_{R}$ theory if these terms
are implemented as ``ad hoc''
constraints on the solutions of the ${\mathcal L}_{2R}$ theory.
This restricted set of solutions is depicted in the bottom shadowed
area in fig.~\ref{fig}.

\avall

\section{Conclusions}
\label{concl}

In this paper we have clarified the issue
of {\sl consistent truncations}, giving them a proper definition
and classifying them into two types: a first-type, corresponding to
a Kaluza-Klein dimensional reduction that keeps unchanged the
number of degrees of freedom attached to every space-time point,
and a second-type, where configuration-space constraints are
introduced that reduce the number of degrees of freedom per
space-time point.
As a byproduct we link in a natural way the issue of consistent
truncations of theories
with that of correct uplifting of solutions.

As regards the first-type, we prove the tracelessness condition
for a consistent truncation under a Lie algebra of {\sl
independent} Killing vector fields. Our proof is at the level of
the Euler-Lagrange equations of motion, and is complementary to
the results in \cite{Scherk:1979zr} and in \cite{Maccallum:gd}. We
emphasise that the proof is complete as regards the aspects of
necessity and sufficiency of the tracelessness condition.
Establishing the conditions for a consistent truncation under Lie
algebras whose generators are not independent remains an open
problem and, up to now, it seems that there is no
group-theoretical argument that can anticipate whether a given
reduction will eventually be a consistent one.

We also discuss the full reduction of the gauge group in the
common cases of the Maxwell $U(1)$ gauge group and the
space-time diffeomorphisms group. The reduction is worked out
in full detail and it is shown in particular
that a residual rigid group of symmetries, associated with
the outer automorphisms of the Lie algebra, remains in the
formalism.

With regard to second-type truncations we show that there is a
useful link with the Dirac-Bergmann theory of constrained systems,
which we find worth to exploit. In particular we explicitate
with an example the fact that the these truncations will
usually be inconsistent due to the presence of secondary
constraints, which are dynamically derived from the primary
ones used to define the truncation. In spite of these
obstructions to a consistent truncation, we show how one can still
have correct upliftings if it happens that there exists a subset
of solutions of the truncated theory that satisfy the secondary
constraints.

Our results indeed reinforce the definition of \emph{consistent truncation}
we make use throughout the paper, reflected in the diagram of the introductory section.

\vskip 6mm

{\it{\bf Acknowledgements}}
J.\ M.\ P.\ thanks the theoretical physics group at the
Imperial College London
for the warm hospitality during the early stages of this work.
This work is partially supported by MCYT FPA, 2001-3598, CIRIT, GC
2001SGR-00065, and HPRN-CT-2000-00131.

\vskip 4mm

\appendix
\setcounter{equation}{0}
\newcounter{zahler}
\addtocounter{zahler}{1}
\renewcommand{\thesection}{\Alph{zahler}}
\renewcommand{\theequation}{\Alph{zahler}.\arabic{equation}}
\label{appdx}

\setcounter{section}{0}
\setcounter{subsection}{0}

\renewcommand{\thesection}{\Alph{zahler}}
\renewcommand{\theequation}{\Alph{zahler}.\arabic{equation}}

\section{Lagrangian $y$-dependences}
\label{proof}

\noindent {\bf Proposition:} All the $y$-dependences in the
Lagrangian are encoded in the single factor $\vert\omega\vert$ in
\bref{factor}.

\vspace{4mm}

\noindent \begin{pf} Since the Lagrangian is a scalar density and all the
fields present in it satisfy the Killing conditions (which is the
point of departure for the reduction procedure), so it does the
Lagrangian itself,
$$
{\mathfrak L}_{{\bf K}_a}{\mathcal L} = \partial_{\alpha}
({\mathcal L}\, K_a^{\alpha}) = 0\,.
$$
On the other hand, the commutation relations (\ref{invarbasis})
imply
\begin{equation}
\partial_{\alpha}( K_a^{\alpha}) =
\vert\omega\vert{\bf K}_a\left(\frac{1}{\vert\omega\vert}\right)\,.
\end{equation}
Then,
\begin{equation}
\partial_{\alpha}\,({\mathcal L}\, K_a^{\alpha}) ={\bf K}_a {\mathcal L} +
{\mathcal L}\vert\omega\vert{\bf K}_a\left(\frac{1}{\vert\omega\vert}\right)
= {\vert\omega\vert}{\bf K}_a
\left(\frac{{\mathcal L}}{\vert\omega\vert}\right)\,,
\end{equation}
together with the condition $ {\mathfrak L}_{{\bf K}_a}{\mathcal L} = 0 $
amounts to ${\bf
K}_a \left(\frac{{\mathcal L}}{\vert\omega\vert}\right)=0$ or, equivalently to,
$\partial_{\alpha}\left(\frac{{\mathcal L}}{\vert\omega\vert}\right)=0$,
which means that ${\mathcal L}$ is
of the form
\begin{equation}
{\mathcal L}(x,y) =\vert\omega\vert f(x)\,,
\end{equation}
for some scalar function $f$ that depends exclusively of the
$x$-coordinates.\end{pf}

\setcounter{equation}{0}
\addtocounter{zahler}{1}
\renewcommand{\thesection}{\Alph{zahler}}
\renewcommand{\theequation}{\Alph{zahler}.\arabic{equation}}

\section{Scalars associated with p-forms and bi-invariance}
\label{scalars}

In this appendix, we shall see that, for semi-simple Lie algebras,
the constraints imposed by the bi-invariance requirement on any
$p$-form, set the components that become scalars under the
reduction process identically to zero.

\vspace{4mm}

\begin{itemize}
\item[{\it{i})}]
In the one-forms case the constraint
 \bref{2fase-onef-cond} gives\footnote{Indices will be raised
 and lowered hereafter using $h_{ab}$.}
$$
0 = \Omega_a C^a_{cd} = \Omega^e h_{ea}C^a_{cd} = \Omega^e
h_{da}C^a_{ce}\, \Rightarrow\, \Omega^e C^a_{ce} = 0\,
\Rightarrow\, \Omega^e h_{ef} = 0\, \Rightarrow\, \Omega^e = 0
$$
which implies
\begin{equation}
\Omega_a = 0\,. \label{oneform0}
\end{equation}

\vspace{4mm}

\item[{\it{ii})}]
When two-forms \bref{twoform} are present, the constraint
\bref{2fase-twof-cond} imposes
$$
0=\Omega_{ba}C^a_{cd} - \Omega_{da}C^a_{cb} = \Omega_{b}^{\ e}
h_{ea}C^a_{cd}- \Omega_{d}^{\ e} h_{ea}C^a_{cb} = \Omega_{b}^{\ e}
h_{da}C^a_{ce}- \Omega_{d}^{\ e} h_{ba}C^a_{ce}\,,
$$
where $h_{ab}$ satisfies \bref{h-cond}. The previous expression
implies, saturating with $h^{bf}h^{dg}$,
\begin{equation}
\Omega^{fe}C^g_{ce} - \Omega^{ge}C^f_{ce} = 0\,. \label{omega-up}
\end{equation}
Saturating \bref{omega-up} with $C^c_{gd}$, one obtains
$$
 \Omega^{f}_{\ d} = \Omega^{ge}C^f_{ce} C^c_{gd} =
\Omega^{ge}(C^f_{cg} C^c_{ed}+C^f_{cd} C^c_{ge})\,.
$$
A direct consequence of \bref{omega-up} is
obtained using the tracelessness condition $C^f_{fe}=0$
\begin{equation}
\Omega^{fe}C^g_{fe}  = 0\,.
\label{omega-up2}
\end{equation}
Using \bref{omega-up2}, the antisymmetry property of $\Omega^{ge}$
and the Jacobi identity we obtain
$$
\Omega^{f}_{\ d} =\Omega^{ge}C^f_{cg} C^c_{ed} = \frac{1}{2}
\Omega^{ge}(C^f_{cg} C^c_{ed}-C^f_{ce} C^c_{gd})=
\frac{1}{2}\Omega^{ge}C^f_{cd} C^c_{ge} = 0\,.
$$
Concluding that
\begin{equation}
\Omega_{ab}  = 0. \label{twof0}
\end{equation}
\end{itemize}

The proof of the vanishing of the scalars coming from p-forms can
be generalised to $p > 2 $. Summing up, our constraints imply
--when the Lie algebra is semi-simple-- that the scalars produced
from the forms must vanish.

\end{document}